\pdfoutput=1

\documentclass{aa}

\usepackage{color}
\usepackage{multirow}
\usepackage{graphicx}
\usepackage{natbib}
\usepackage{epstopdf}
\usepackage[varg]{txfonts}

\begin{document}

   \title{WASP\,0639-32: a new F-type subgiant/K-type main-sequence detached eclipsing binary from the WASP project}


   \author{J.~A.~Kirkby-Kent\inst{1}
          \and
          P.~F.~L.~Maxted\inst{1}
          \and
         {A.~M.~Serenelli}\inst{2}
          \and
          D.~R.~Anderson\inst{1}
          \and
          C.~Hellier\inst{1}
         \and
         R.~G.~West\inst{3}
          }

   \institute{Astrophysics Group,  Keele University, Keele, Staffordshire ST5
5BG, UK\\
              \email{j.a.kirkby-kent@keele.ac.uk}
 	\and 
	Institute of Space Sciences (ICE/CSIC-IEEC), Carrer de Can Magrans S/N, E-08193, Cerdanyola del Valles, Spain
	\and
	Department of Physics, University of Warwick, Coventry, CV4 7AL, UK}

   \date{Received ; accepted }

 
  \abstract
   {}
   {Our aim is to measure the masses and radii of the stars in a newly-discovered detached eclipsing binary system to a high precision ($\approx$1\%), enabling the system to be used for the calibration of free parameters in stellar evolutionary models.}
   {Photometry from the Wide Angle Search for Planets (WASP) project was used to identify \object{1SWASP J063930.33-322404.8} (\object{TYC 7091-888-1}, WASP\,0369-32 hereafter) as a detached eclipsing binary system with total eclipses and an orbital period of $P=11.66$ days. Lightcurve parameters for WASP\,0639-32 are obtained using the {\textsc{ebop}} lightcurve model, with standard errors evaluated using a prayer-bead algorithm. Radial velocities were measured from 11 high-resolution spectra using a broadening function approach, and an orbit was fitted using {\textsc{sbop}}. Observed spectra were disentangled and an equivalent width fitting method was used to obtain effective temperatures and metallicities for both stars. A Bayesian framework is used to explore a grid of stellar evolution models, where both helium abundance and mixing length are free to vary, and use observed parameters (mass, density, temperature and metallicity) for each star to obtain the age and constrain the helium abundance of the system.}
   {The masses and radii are found to be $M_{1}=1.1544\pm0.0043\,M_{\sun}$,  $R_{1}=1.833\pm0.023\,R_{\sun}$ and $M_{2}=0.7833\pm0.0028\,M_{\sun}$, $R_{2}=0.7286\pm0.0081\,R_{\sun}$ for the primary and secondary, respectively. The effective temperatures were found to be $T_{1}=6330\pm50$\,K and $T_{2}=5400\pm80$\,K for the primary and secondary star, respectively. The system has an age of $4.2^{+0.8}_{-0.1}$\,Gyr, and a helium abundance in the range 0.251-0.271.}
  {WASP\,0639-32 is a rare example of a well-characterised detached eclipsing binary system containing a star near the main-sequence turn-off point. This makes it possible to measure a precise age for the stars in this binary system and to estimate their helium abundance. Further work is needed to explore how this helium abundance estimate depends on other free parameters in the stellar models.}
   \titlerunning{WASP\,0639-32: A new detached eclipsing binary system}
   \maketitle
%

\section{Introduction}
\label{sec:Intro}

Detached eclipsing binary stars with well-determined parameters are one of the best methods for testing stellar evolutionary models \citep{2002A&A...396..551L, 2010A&ARv..18...67T}. These models play an important role in areas such as exoplanet research, where they are used to determine the mass and age of planet-host stars, and in galaxy evolution modelling, where they are used to predict the properties of the stars that are observed by surveys of the Milky Way and other galaxies \citep{2013ASSL..396..345S}.

Stellar evolutionary models can use different prescriptions to describe complex phenomena such as convection and diffusion. The free parameters used in these prescriptions are, in general, poorly constrained by the underlying physics, so they are either calibrated from observations, or by detailed simulations.  Creating models for stars that are different from the stars used in the calibrations, whilst using the same calibrated parameters values, can introduce systematic errors into the models. As such, these free parameters are still large sources of uncertainty in using stellar models to infer the ages and other properties of stars \citep{2014EAS....65...99L}. Recently, research has been pushing to understand these parameters in more detail, with \cite{2017A&A...600A..41V} looking at how uncertainties on observational data affect the determination of these parameters, in particular the overshooting parameter. They note, following on from an earlier paper \citep{2016A&A...587A..16V}, that for binaries where both stars are on the main-sequence, the calibration of the overshooting parameter is hampered by uncertainties of the order 0.5\% in the radii and 1\% in the masses. This indicates that in order to calibrate the overshooting, binary systems with stars at two different evolutionary stages must be used and their parameters must be known to very high precision, \citep[at least 1\% in the masses,][]{2017A&A...600A..41V}. The number of well-studied binary systems with both the required precision, and the required evolutionary state are very few. Excluding AI Phe, and TZ For (see below) there is only one other system in the list by \cite{2010A&ARv..18...67T} that meets these requirements, and can be classed as a solar-type star, that is {\object{V432 Aur}}. This system also has a relatively short 3-day orbital period, meaning there is a strong possibility that tidal interactions between the stars would invalidate the assumption that the two stars have evolved independently. Some examples which meet the requirements include {\object{LL Aqr}} \citep{2016A&A...594A..92G}, {\object{TZ For}} \citep{2017A&A...600A..41V}, and two binaries in {\object{NGC 6791}} \citep{2012A&A...543A.106B}. Each have provided a constraint on the helium abundance of the particular binary system. \cite{2016A&A...591A.124K} looked at the binary system {\object{AI Phe}}, and after improving the mass and radius estimates of both stars, looked at how independently varying the mixing length and helium abundance affected the age of the system. Again the precise observational constraints allowed the helium abundance $Y_{\rm i}$  to be constrained to the range 0.25-0.28.

If meaningful statistics are to be done on the helium abundance in low-mass stars, then more systems need to be studied to the same level of precision. This paper details the photometric and spectroscopic analysis of a newly-discovered detached eclipsing binary system containing an evolved F-type primary and a main-sequence K-type star, which was first identified using data from the Wide Angle Search for Planets \citep[WASP,][]{2006PASP..118.1407P}. Section \ref{sec:observations} describes the data that used in the photometric analysis (Sect. \ref{sec:PhotometryAnalysis}) and the spectra used in the orbital analysis (Sect. \ref{sec:RVAnalysis}) and the spectroscopic analysis (Sect. \ref{sec:specAnalysis}). Section \ref{sec:massAndRadii} combines the lightcurve and spectroscopic orbit parameters to give masses and radii of the system and in Sect. \ref{sec:models} we investigate how the helium abundance can be constrained by the precise binary parameters system. Finally, discussion and conclusion sections follow in Sects. \ref{sec:Discuss} and \ref{sec:Conc}.

\section{Observations}
\label{sec:observations}

\subsection{WASP photometry} 
\label{subsec:WASPphotometry}

The WASP project \citep{2006PASP..118.1407P} uses two instruments, one located at the Observatorio del Roque de los Muchachos, La Palma and the other at Sutherland Observatory, South Africa. Each instrument consists of eight wide-field cameras each with a 2048 x 2048 pixel CCD. In total, 39\,621 photometric measurements were obtained for WASP\,0639-32 (\object{1SWASP J063930.33-322404.8}, \object{TYC 7091-888-1}) between May 2006 and March 2012 by the WASP-South instrument using 200-mm, f/1.8 lenses and broadband 400-700\,nm filters \citep{2006PASP..118.1407P}. A specialised pipeline \citep{2006PASP..118.1407P} is used to reduce the images and the data are subject to a detrending algorithm described by \citet{2006MNRAS.373..799C}.

\subsubsection{Initial Processing}
\label{subsubsec:InitialProcessing}

The WASP data can suffer from significant amounts of scatter caused by clouds, scattered moonlight, etc. To remove the affected observations, the photometric data is process by an algorithm, that was first described in \cite{2016A&A...591A.124K}. We briefly discuss the process here.

Firstly, each measurement that has the weighting factor, $\sigma_{\rm xs}$, set to zero, is removed. This factor is defined as part of the WASP pipeline \citep[$\sigma_{t(i)}$ in][]{2006MNRAS.373..799C} and characterises the amount of scatter present from external sources. When set to zero, it indicates the pipeline as has marked the value as missing or bad. Data where the uncertainty in a flux measurement was more then five times the median value were also removed. One WASP field contained only ten observations, which were removed as they caused problems when the analysis compared data from different nights and seasons. 

The WASP pipeline provides the flux, $f$, of WASP\,0639-32 measured using an aperture of 3.5 pixels, and also the uncertainty due to known noise sources (photon-counting noise, background subtraction, etc.), $f_{\rm err}$. Trends in the data are removed using a detrending algorithm \citep{2006MNRAS.373..799C}, which is described in more detail in Sect. \ref{subsec:Detrend}. The flux measurements are converted into magnitudes, using the overall median flux as the zero point, and the uncertainties, $m_{\rm err}$, are calculated using
\begin{equation}
\mbox{$m_{\rm err}$} = f \sqrt{\left(\frac{f_{\rm err}}{f}\right)^{2} + {\sigma_{\rm xs}}^2}
\label{eq:error}
\end{equation}
The final step looks for data that is significantly offset from the other data. This technique uses a model generated from 1\,000 bins across the phase-folded data, with the median magnitude in each bin used as a reference. The magnitude from each observation within a block of data (the observation from one camera on one night) was compared to the expected value within in a phase bin. If observations differed by more than ten times the uncertainties for each phase bin, it was removed, and if more than 80\% percent of data from a block was offset, the entire block was removed.

Overall, from the initial 39\,621 observations stored in the WASP archive, 28\,566 remained for use during the analysis. Table~\ref{tab:AllPointsRemoved} provides a summary of the number of points removed during this initial processing stage. 

\begin{table}
\caption{Number of observations removed during initial processing.}
\label{tab:AllPointsRemoved}
\centering
\begin{tabular}{l r}
\hline\hline
\noalign{\smallskip}
Reason				& Points removed\\
\noalign{\smallskip}
\hline
\noalign{\smallskip}
$\sigma_{\rm xs} = 0$	& $4\,870$		\\
$f_{\rm err} >$ 5*median	& $2\,408$		\\
Small field				& $10$		\\
Offset				& $3\,766$		\\ 
\noalign{\smallskip}
\hline
\noalign{\smallskip}
Remaining			& $28\,566$\\
\noalign{\smallskip} 
\hline
\end{tabular}
\tablefoot{Last row states the number of observations remaining.}
\end{table}

\subsection{UVES spectra} 
\label{subsec:UVESspectra}

The eleven spectra used in this work, were obtained using the Ultraviolet and Visual \'Echelle Spectrograph \citep[UVES,][]{2000SPIE.4008..534D} at the Very Large Telescope (VLT), over a period between 9th October 2014 and 20th January 2015. A wavelength range of approximately 500-700\,nm, with a small gap at 600\,nm, and a signal-to-noise ratio (SNR) of over 100 was obtained for each spectrum. The slit was set to a width of 0.7 arcseconds, resulting in spectra with a resolving power $R=56990$. Each spectrum was reduced using the standard UVES reduction pipeline \citep{2000Msngr.101...31B}.

\section{Photometric Analysis}
\label{sec:PhotometryAnalysis}

\subsection{Ephemeris}
\label{subsec:ephemeris}

\begin{figure}
\resizebox{\hsize}{!}{\includegraphics{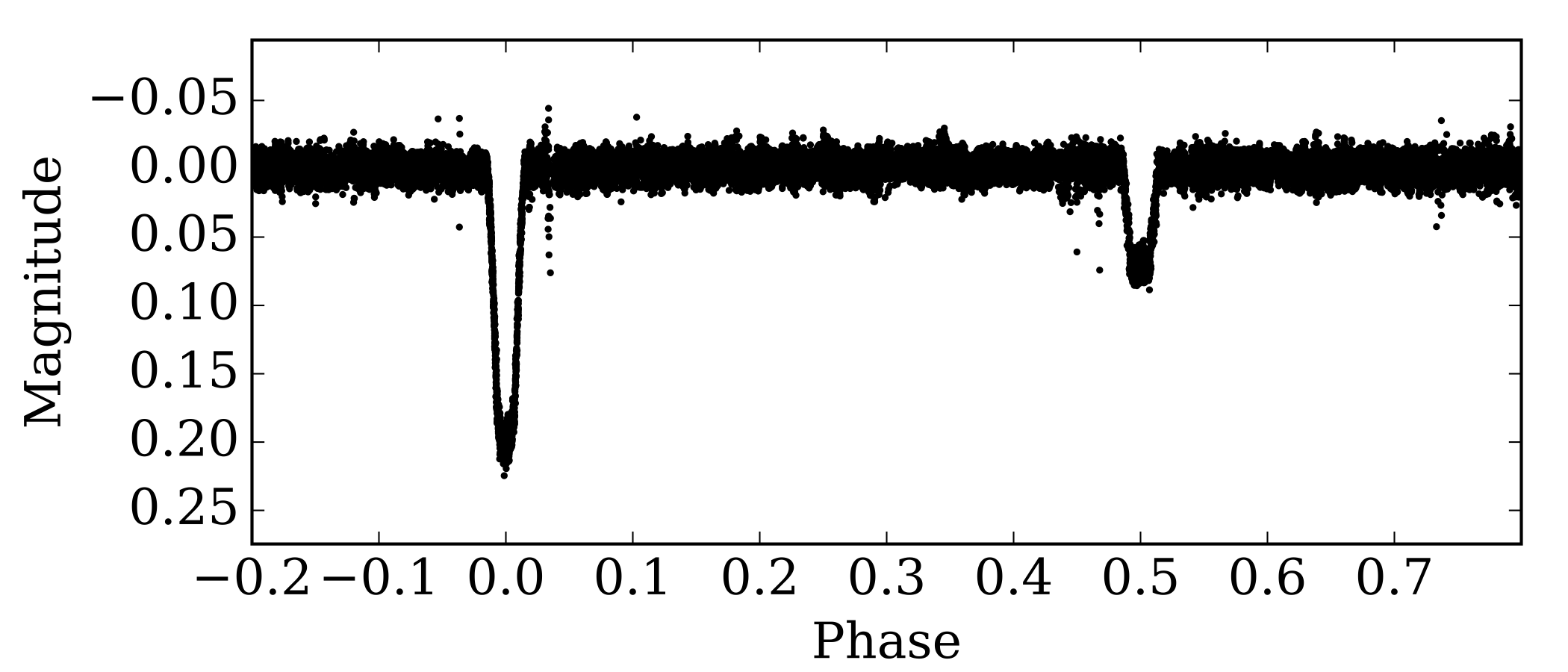}}
\caption{Phase-folded lightcurve for WASP\,0639-32, showing data used for lightcurve analysis.} 
\label{fig:phaseFoldedLC}
\end{figure}

To find the period, $P$, and time of primary minimum, the WASP photometry, with very minimal cleaning applied (i.e. only the points with $\sigma_{\rm xs}=0$ where removed), was fitted using \textsc{jktebop} \citep{JKTebopRef}. The surface brightness ratio, sum of radii, ratio of radii, orbital inclination, $e\cos \omega$ and $e\sin \omega$ parameters were included in the model fitting. The results for the period and time of primary minimum are shown in Eq. (\ref{eq:ephemeris}).
\begin{equation}
\mbox{\rm HJD Pri. Min.} = 2\,455\,241.33931(26) + 11.658317(5)\,E.
\label{eq:ephemeris}
\end{equation}
The period and time of primary minimum have been kept fixed in all subsequent analysis.  
Figure~\ref{fig:phaseFoldedLC} shows the photometric measurements phase-folded using the ephemeris in Eq. (\ref{eq:ephemeris}), after the processing described in Sect. \ref{subsubsec:InitialProcessing}.

\subsection{Lightcurve Modelling}
\label{subsec:Model}

To model the lightcurve we used the same method as described in \citet{2016A&A...591A.124K}. It uses the subroutine {\tt{light}} from the \textsc{ebop} lightcurve analysis code \citep{1972ApJ...174..617N, 1981AJ.....86..102P}, with modifications allowing it to be called directly from the Python programming language. The parameters used in the \textsc{ebop} model are optimised using the least-squares Levenberg-Marquardt Python module, \textsc{MPFIT} \citep{2009ASPC..411..251M}.

The model has seven parameters that were fitted: surface brightness ratio at the centre of the stellar discs, $J$; sum of the radii, $r_{\rm sum} = r_{1} + r_{2}$; ratio of the radii, $k = r_{2}/r_{1}$; inclination, $i$; $e\cos{\omega}$, $e\sin{\omega}$ and third-light, $l_{3}$. In addition to these fitted parameters, the fractional radii, $r_{1}$ and $r_{2}$, are automatically calculated from $r_{\rm sum}$ and $k$, while the eccentricity, $e$, and longitude of periastron, $\omega$, are calculated from $e\cos{\omega}$ and $e\sin{\omega}$.

The mass ratio, $q = M_{2}/M_{1}$, was fixed at 0.67. This value was determined by initially setting $q=0.5$, fitting a model and then using the estimates of $r_{1}$, $r_{2}$ and the spectroscopic orbit parameters (see Sect. \ref{sec:RVAnalysis}) as inputs into {\textsc{jktabsdim}}\footnote{\tt{http://www.astro.keele.ac.uk/jkt/codes/jktabsdim. html}}. For WASP\,0639-32, the mass ratio contributes very little to the overall shape of the lightcurves meaning variations of 0.01 will not alter the final best-fit parameters. The gravity darkening exponents were fixed at 0.26 and 0.46 for the primary and secondary component, respectively \citep{2011A&A...529A..75C}. These parameters also have very little effect on the lightcurve as the stars are nearly spherical in shape.

Attempts to include the limb darkening coefficients as free parameters in the fitting proved unsuccessful as these coefficients are not well-constrained by the data. Instead, the coefficients were fixed at values taken from interpolating between values in the tables by \citet{2011A&A...529A..75C}. The {\em{Kepler}} passband was used to approximate the response of the WASP broadband filter. The adopted primary limb-darkening coefficient was $u_{\rm p}=0.50\pm0.05$ and $u_{\rm s}=0.63\pm0.03$ was adopted for the secondary. To account for the uncertainties in the limb-darkening coefficients, models were calculated with the fixed limb darkening parameters varied by their uncertainties. The average scatter in the parameters from these models has been added in quadrature to the uncertainties from the best-fit model. Table \ref{tab:LDuncert} details the uncertainty in each parameter due to the uncertainty in the limb-darkening coefficients.

\begin{table}
\caption{Uncertainty contribution to each parameter from uncertainty in the limb darkening coefficients.}
\label{tab:LDuncert}
\centering{
\begin{tabular}{l r r}
\hline\hline
\noalign{\smallskip}
Parameter			&	Original 	& Detrend \\
\noalign{\smallskip}
\hline
\noalign{\smallskip}
$J$				& $0.0042$	& $0.0042$	 \\
$r_{\rm sum}$		& $0.00006$	& $0.00003$	   \\
$k$ 				& $0.003$	 	& $0.003$		   \\
$i$ (\degr)			& $0.005$		& $0.005$		   \\
$e \cos \omega$ 	& $0.000003$ 	& $0.000005$	  \\
$e \sin \omega$ 	& $0.002$ 	& $0.002$	  \\
$l_{3}$			& $0.005$		& $0.005$		  \\
\noalign{\smallskip}
\hline
\noalign{\smallskip}
$r_1$ 			& $0.00013$ 	& $0.00013$ 	  \\
$r_2$			& $0.00013$	& $0.00016$	   \\
$e$				& $0.0020$	& $0.0019$		\\
$\omega$ (\degr)	& $0.7$		& $0.8$		 \\
\noalign{\smallskip} 
\hline
\end{tabular}}
\end{table}

\begin{figure}
\resizebox{\hsize}{!}{\includegraphics{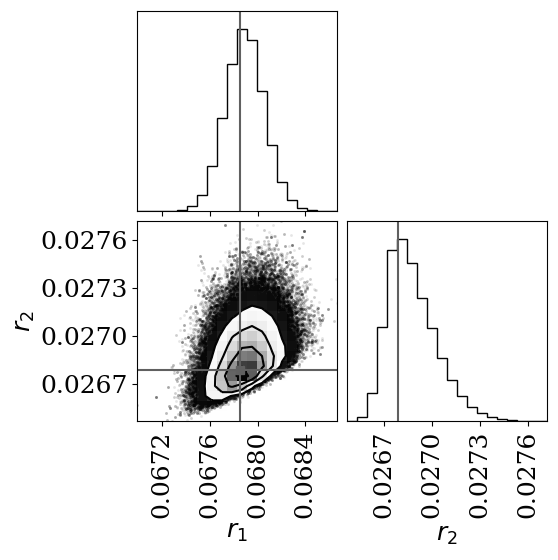}}
\caption{Density distribution of the fractional radii, $r_{1}$ and $r_{2}$, for WASP\,0639-32. Grey cross marks the values calculated from the best-fit values from the fit to the `original' data.}
\label{fig:r1r2Correlation}
\end{figure}

As in \citet{2016A&A...591A.124K}, the parameter-space is explored using Markov Chain Monte Carlo (MCMC), in the form of the Python module, \texttt{emcee}, \citep{2013PASP..125..306F}. It was used to ensure the lightcurve solution was not a local minimum in the parameter space. {\texttt{emcee}} uses a affine-invariant ensemble sampling (stretch-move) algorithm of \cite{GoodmanWeare2010}. By using affine-invariant transformations, the algorithm can work with skewed distributions. The algorithm explores the parameter space using a group of walkers, which can be split allowing the process to be run in parallel. Walker positions in a particular sub-group will be updated using the positions of walkers in other sub-groups. The probability that a model produced by a set of parameters, corresponds to the best-fit model, is evaluated using the log-likelihood function
\begin{equation}
\ln \mathcal{L}({\bf y }; {\bf \Theta}) = -\frac{1}{2} \sum^N_{n=1} \left[ \left(\frac{m_{n}-y_{n}({\bf \Theta})}{m_{{\rm {err,}}n}}\right)^2 - \ln \left( \frac{2\pi}{m_{{\rm{ err,}}n}^2}\right) \right]
\label{eq:loglike}
\end{equation}
where {\bf y} is a vector of length $N$ containing the magnitudes generated for a model, {\bf $\Theta$} is a vector containing the varying parameters ($J$, $r_{\rm sum}$, $k$, $i$, $e\cos{\omega}$, $e\sin{\omega}$ and $l_3$), $m$ is the observed magnitude and $m_{\rm err}$ is the standard error on the magnitude. Priors were applied, but these are only used to prevent the parameter exploring areas that are unphysical, e.g. $r_{\rm sum}$ or $J$ being less than zero. The MCMC process ran using 150 walkers for 2\,500 steps, of which the first 200 were discarded to allow for an adequate burn-in stage. For each of the walkers, a starting point for each parameter was chosen by choosing a number at random from a normal distribution (with a mean of zero and variance of 0.01) and adding it to the best-fit parameter. This method creates a ``ball'' of walkers close to the best-fit solution, however during the burn-in stage, the walkers spread out from this ball. The chains for each parameter were checked to ensure the burn-in stage was completed within these first 200 steps and adequate mixing had occurred. We also check for bias in the starting positions by carrying out a run where the walkers' starting positions can be distributed by up to three times the uncertainty on the parameters, as determined by the covariance matrix obtained from the best fit. The burn-in stage for this test was longer, but still produced the same distribution. Various checks were performed to ensure suitable mixing and convergence. The walkers' positions were plotted against step number to visually check the mixing and to determine the burn-in length. Convergence is checked through a running-mean test. We also check the acceptance fraction (mean$=0.43$) and auto-correlation length (max$=92$) for each run. The auto-correlation length is a measure of the number of evaluations needed to have independent samples \citep{2013PASP..125..306F}. Parameter uncertainties were calculated using a prayer-bead method, again as described in \citet{2016A&A...591A.124K}. The prayer-bead algorithm used 500 shifts, spread evenly across the data, with starting positions selected randomly from the MCMC analysis. The prayer-bead algorithm allows robust estimates of the uncertainties associated with lightcurve parameters (as the standard deviation of the best-fit values from the 500 shifts) and accounts for the correlated noise that can be found within the WASP data, and varying the starting positions overcomes any bias that may be present by the choice of initial parameters. Figure \ref{fig:r1r2Correlation}\footnote{Produced using {\texttt{corner.py} by \citep{corner}.}} shows the density distribution for $r_{1}$ and $r_{2}$, which has been calculated from the $k$ and $r_{\rm sum}$ positions explored during the MCMC. The cut-off in the distribution occurs where the third-light becomes zero, as there strong correlations between $k$ and $l_{3}$. The parameters $k$ and $r_{\rm sum}$ are directly related to observable features in the lightcurve and so there values are well determined.

\subsection{Detrending Investigations}
\label{subsec:Detrend}

\begin{table}
\caption{Best-fit parameters for WASP\,0639-32, with and without the detrending applied. }
\label{tab:DetrendData}
\centering{
\begin{tabular}{l r r r}
\hline\hline
\noalign{\smallskip}
Parameter			& Original		& Detrended	 & Difference	\\
\noalign{\smallskip}
\hline
\noalign{\smallskip}
$J$				& $0.4527(72)$		& $0.4513(71)$		& $0.0014$	 \\
$r_{\rm sum}$		& $0.09465(42)$	& $0.09499(42)$	& $-0.00034$	   \\
$k$ 				& $0.3949(53)$	 	& $0.3975(53)$		& $-0.0026$ 	     \\
$i$ (\degr)			& $89.9925(68)$	& $89.9943(67)$	& $-0.0018$	    \\
$e \cos \omega$ 	& $-0.00027(12)$ 	& $-0.00026(12)$	& $-0.00001$		  \\
$e \sin \omega$ 	& $-0.0077(44)$ 	& $-0.0070(44)$	& $-0.0007$ 		  \\
$l_{3}$			& $0.007(20)$		& $0.012(20)$		& $-0.005$	 \\
\noalign{\smallskip}
\hline
\noalign{\smallskip}
$r_1$ 			& $0.06785(86)$ 	& $0.06797(86)$ 	& $-0.00012$	 \\
$r_2$			& $0.02679(29)$	& $0.02702(30)$	& $-0.00023$	   \\
$e$				& $0.0077(54)$ 	& $0.0070(53)$		& $0.0007$	\\
$\omega$ (\degr)	& $268.0(4.8)$ 		& $267.9(4.9)$		& $0.1$		 \\
\noalign{\smallskip}
\hline
\end{tabular}}
{\tablefoot {Standard errors on the final two digits of each parameter value are given in the parentheses and include the contribution from the uncertainties in the limb darkening coefficients.}}
\end{table}

To help ensure the most accurate parameters possible were obtained from the lightcurve, we have investigated whether the WASP detrending algorithm alters the best-fit parameters for WASP\,0639-32. The algorithm locates and corrects for four trends of systematic errors found across all stars in one field. It can be described using
\begin{equation}
\widetilde{m}_{i, j} = m_{i,j} - \sum^{M}_{k=1} {^{(k)}}c_{j} {^{(k)}}a_{i}
\label{eq:detrendCC}
\end{equation}
where $m_{i,j}$ and $\widetilde{m}_{i,j}$ are the observed and corrected magnitude, respectively, for star $j$ at time $i$. $M$ is the total number of trends, $a_{i}$ are basis functions detailing the patterns of systematic errors and $c_{j}$ describes to what extent each basis function affects a particular star.

Using the method described in \citet{2016A&A...591A.124K}, effective detrending coefficients, $c{'}$, were calculated using singular value decomposition (SVD). The effective detrending coefficients include a fixed binary lightcurve model, $L$, when they are calculated to allow the variability of an eclipsing to be taken into consideration. It can be described using
\begin{equation}
\widetilde{m}_{i} = m_{i} + L_{i} - \sum^{M}_{k=1} {^{(k)}}c{'} {^{(k)}}a_{i} 
\label{eq:detrend}
\end{equation}
where $m_{i}$ and $\widetilde{m}_{i}$ are the observed and corrected magnitude (respectively), and $ {^{(k)}}a_{i}$ are the same detrending basis functions as in Eq. (\ref{eq:detrendCC}). Once calculated, the trends are removed from the observed data and a new model fitted. Initially the best-fit model from Sect. \ref{subsec:Model} was used for $L$, but the coefficients were calculated in an iterative process with slightly different models until each parameter change by less than 0.005\% from the previous model. As before, the same MCMC and prayer-bead analysis was used as in Sect. \ref{subsec:Model}.

\begin{figure}
\centering
\resizebox{\hsize}{!}{\includegraphics{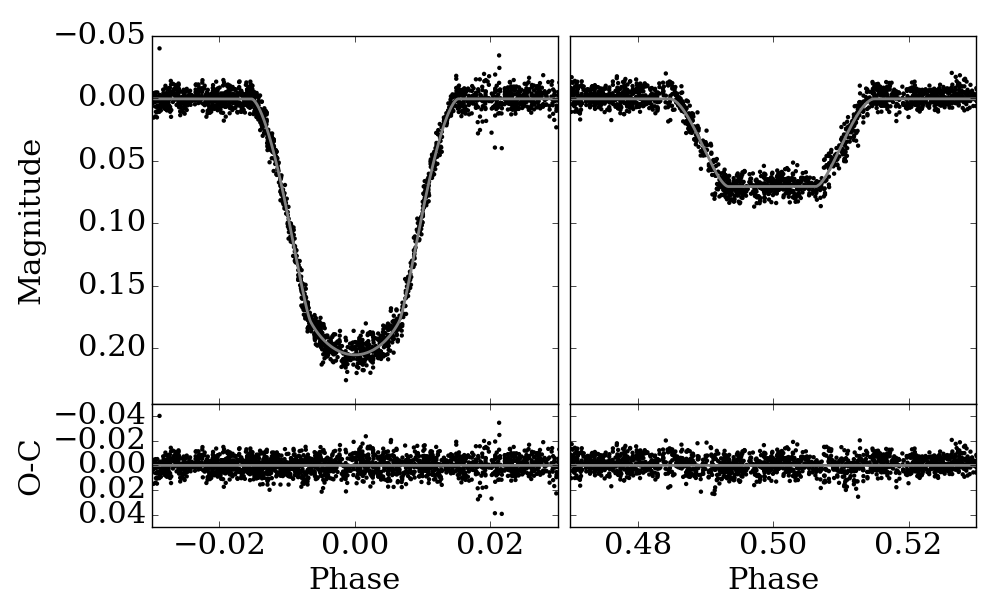}}
\caption{Upper panels - the detrended best-fit model for WASP\,0639-32 (grey line) plotted over the WASP-South photometry for the primary (left) and secondary (right) eclipses. Lower panels - the residuals, with the grey line marking zero.} 
\label{fig:bestFit}
\end{figure}

Table \ref{tab:DetrendData} contains best-fit parameters for both the original model fit and those from the detrending testing. The difference between the two is also shown for ease of comparison. In all cases the difference is less that the uncertainties associated with that parameters, so the WASP detrending algorithm does not affect the lightcurve parameters for WASP\,0639-32. This is the same conclusion as was found the analysis of AI Phe \citep{2016A&A...591A.124K}. The detrended parameters have been used in the remaining analysis. Figure \ref{fig:bestFit} shows the detrended best-fit model for WASP\,0639-32 plotted against the WASP-South photometry.

\section{Spectroscopic Orbit}
\label{sec:RVAnalysis}

Radial velocity measurements for both stars were made using the broadening function method \citep{2002AJ....124.1746R,1999AJ....118.2451R,1992AJ....104.1968R} in RaveSpan \citep{2017ApJ...842..110P}. The broadening function smoothing was set to a value of 3 and a template spectrum with parameters of T$_{\rm eff}=6200\,K$, $\log g=4.0$, [Fe/H]$=0.0$ and $v \sin (i)=0.0$ km\,s$^{-1}$ was used. The measured radial velocities are shown in Table \ref{tab:j0639RVs}. The spectra were obtained on separate nights to avoid correlations between measured radial velocities.

\begin{table} 
\caption{Radial velocities for WASP\,0639-32 measured using broadening function method within RaveSpan.}
\label{tab:j0639RVs}
\centering{
\begin{tabular}{l r r}
\hline\hline
\noalign{\smallskip}
HJD$-2\,450\,000$			& Primary 		& Secondary  \\
						& (km\,s$^{-1}$) &(km\,s$^{-1}$) \\
\noalign{\smallskip}
\hline
\noalign{\smallskip}
6939.78595				& $100.90$	& $-5.61$		 	\\
6939.85977				& $101.44$	& $-6.86$		\\
6940.81320				& $104.62$	& $-10.61$			\\
6956.80762				& $20.29	$	& $113.83$		\\
6958.81865				& $14.74$		& $122.02$		 \\
6985.78700				& $92.35$		& $7.10$ 	 \\
6999.65212				& $100.42$	& $-4.35$		\\
6999.77011 				& $98.29$		& $-3.15$		\\
7015.77003				& $12.02$		& $124.57$	\\
7040.65621				& $16.67$		& $117.65$	\\
7042.59910\parbox{0pt}{$^*$}	& $58.74$		& ...	\\
\noalign{\smallskip}
\hline
\end{tabular}}
\tablefoot{ Typical internal errors: primary, 0.03\,km\,s$^{-1}$; secondary, 0.26\,km\,s$^{-1}$. \tablefoottext{*}{Taken at phase $\phi=0.504$, during the flat section of secondary eclipse.}}
\end{table}

Orbital parameters were fitted using {\textsc{sbop}}, the Spectroscopic Binary Orbit Program (written by P. B. Etzel), with an MCMC wrapper in the form of the Python module \texttt{emcee}, \citep{2013PASP..125..306F} to ensure robust uncertainties on the orbital parameters. The following parameters were included in the fitting: eccentricity, $e$; angle of periastron, $\omega$; systemic velocity, $v_{\rm sys}$; and the semi-amplitude of the primary and secondary, $K_{1}$ and $K_{2}$ respectively. The orbital period and time of periastron were fixed at the period and time of primary minimum from Eq. (\ref{eq:ephemeris}) because the fitted lightcurve parameters showed that the system is very close circular. In circular systems degeneracy means it is not possible to fit the eccentricity, angle of periastron and time of periastron simultaneously.

Initial fits showed an average offset of $0.56$\,km\,s$^{-1}$ between the observed and computed (O-C) measurements for the secondary star. As such, a constant offset parameter, $B_{0}$, was fitted alongside the orbital parameters. It is likely that this offset is due the large difference in the spectral types of the two stars, having focused the template on parameters for the hotter, primary star. The offset will also, in part, be due the effects of differing gravitational redshifts and convective blueshifts between the two stars \citep{2003A&A...401.1185L}.

For one component, $j$, the log-likelihood can be written as
\begin{equation}
\label{eq:rvloglike}
\ln \mathcal{L}_{\rm j}({\bf y}_{\rm rv};{\bf \Theta}_{\rm rv}) = -\frac{1}{2}
\left[ \sum^N_{n=1} \left(\frac{r_{{\rm j},n}-y_{{\rm j,}n}({\bf \Theta}_{\rm rv})}{s_{{\rm j,} n}^{2}} \right)^2 
- \ln \left( \frac{2\pi}{s_{{\rm j,} n}^2}\right) \right]
\end{equation}
where {\bf{y}}$_{\rm rv}$ is a vector of length $N$ containing the modelled radial velocities of star $j$, {\bf $\Theta$}$_{\rm rv}$ is a vector containing the varying parameters ($e$, $\omega$, $\gamma_{\rm sys}$, $K_{1}$, $K_{2}$, $B_{0}$, $\sigma_{\rm sys,1}$, $\sigma_{\rm sys,2}$) and $r_{\rm j}$ are measured radial velocities for component $j$. $s_{{\rm j,}n}$ are weights for each measurement that combine the measured internal uncertainties, $\sigma_{{\rm j},n}^{2}$ and a systematic uncertainty $ \sigma_{{\rm sys, j}}^{2}$, such that 
\begin{equation}
\label{eq:rvweights}
s_{j,n}^2 = \sigma_{{\rm{j}},n}^{2} + \sigma_{{\rm sys, j}}^{2}.
\end{equation}
$ \sigma_{{\rm sys, j}}^{2}$ represents a combination of instrumental uncertainty and stellar jitter for component $j$. Stellar jitter is more significant for more evolved stars \citep{2005PASP..117..657W}, therefore $ \sigma_{{\rm sys}}^{2}$ was included separately for each component. The log-likelihood from both components are summed together as
$\ln \mathcal{L} = \ln \mathcal{L}_{\rm 1}+\ln \mathcal{L}_{\rm 2}$
where the labels $j=1,2$ denote the primary and secondary component, respectively. The chosen priors place loose constraints on the chain parameters to prevent the walkers exploring regions of the parameter-space that correspond to non-physical values. Priors on the semi-amplitude velocities are set to allow exploration between $-500$ and $500$\,km\,s$^{-1}$, and the systemic velocity is restricted to between $-200$ and $500$\,km\,s$^{-1}$. The MCMC used 300 walkers and ran for 1000 steps, with the first 400 steps removed as burn-in steps. The chains from each parameter were checked via a running-mean to ensure the chains had converged, and the walkers' positions were plotted against step number to ensure there is suitable mixing. The acceptance fraction (mean{$=0.40$}) and parameter auto-correlation lengths (max{$=87$}) are also checked.

\begin{table} 
\caption{Fitted spectroscopic orbit parameters for WASP\,0639-32.}
\label{tab:j0639orbitParam}
\centering{
\begin{tabular}{l r | l r}
\hline\hline
\noalign{\smallskip}
Parameter				& \multicolumn{1}{c}{Best-fit}  & Parameter & \multicolumn{1}{c}{Best-fit} \\
					&  \multicolumn{1}{c}{value} & &\multicolumn{1}{c}{value} \\
\noalign{\smallskip}
\hline
\noalign{\smallskip}
$K_{1}$ \mbox{(km s$^{\rm -1})$}	& $47.32(8)$	& $\omega$ (\degr) 				& $269.96(12)$	\\
$K_{2}$ \mbox{(km s$^{\rm -1})$}	& $69.74(11)$	& $B_{0}$ \mbox{(km s$^{\rm -1})$}		& $0.85(13)$	\\
$\gamma_{\rm sys}$ \mbox{(km s$^{\rm -1})$}	& $57.46(8)$ & $\sigma_{\rm sys,1}$	\mbox{(km s$^{\rm -1})$}				& $0.26(7)$	\\
$e$							& $0.0009(^{+12}_{-06})$	 & $\sigma_{\rm sys,2}$ \mbox{(km s$^{\rm -1})$}			& $0.09(10)$	\\
\noalign{\smallskip}
\hline
\end{tabular}}
\end{table}

\begin{figure}
\centering
\resizebox{\hsize}{!}{\includegraphics{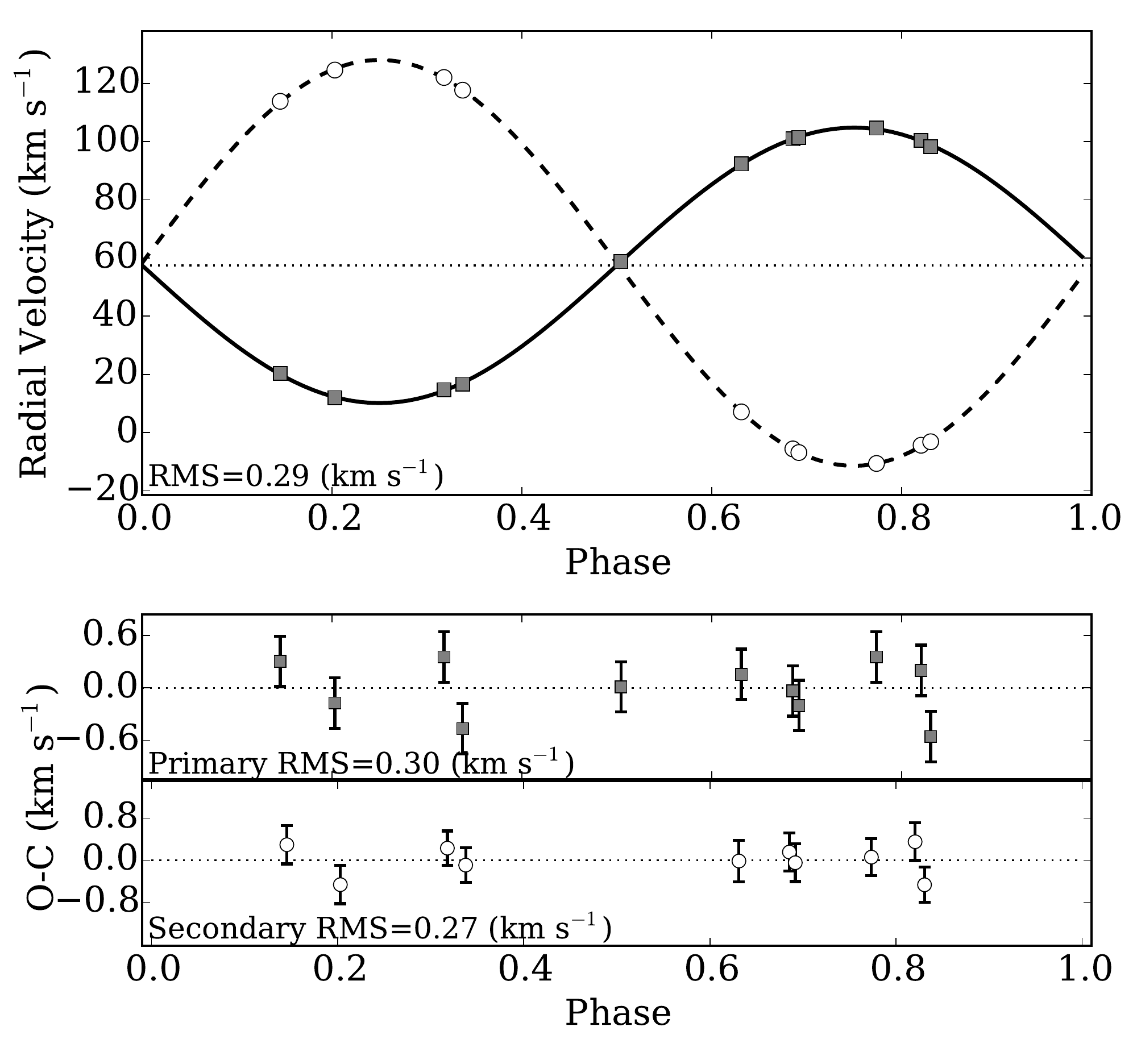}}
\caption{Upper panel - Radial velocities for the primary (grey squares) and secondary (white circles) components of WASP\,0639-32, with best-fit orbit. Lower panels - Residuals for each component. Error bars include the internal and systematic uncertainties.} 
\label{fig:bestFitRVs}
\end{figure}

The resulting orbit parameters are shown in Table \ref{tab:j0639orbitParam} and the resulting model orbits are plotted in Fig. \ref{fig:bestFitRVs}. The parameters and their associated uncertainties are calculated from the 15.9, 50 and 84.1 percentiles. The eccentricity is consistent with a circular orbit, but it has been allowed to vary in order for the uncertainty to be accounted for in $K_{1}$ and $K_{2}$. $K_{1}$ and $K_{2}$ are the key parameters used to determine the masses of the stars. Both are determined directly from the range of the radial velocity curve, so their values are well determined.

\section{Spectroscopic Analysis}
\label{sec:specAnalysis}

\subsection{Disentangling}
\label{subsec:disentangling}

\begin{figure}
\centering
\resizebox{\hsize}{!}{\includegraphics{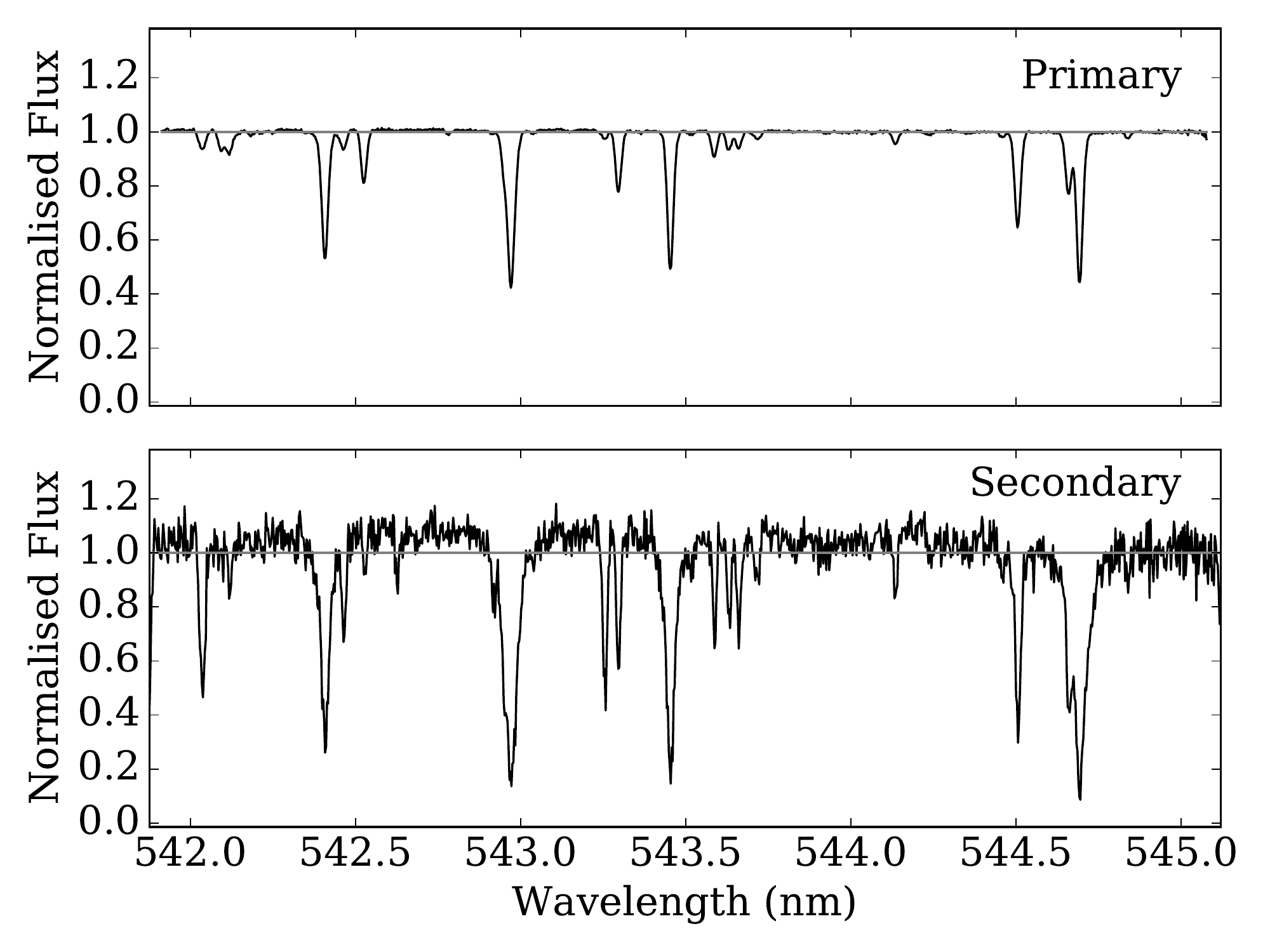}}
\caption{Disentangled spectra for the primary (upper panel) and secondary (lower panel) components in WASP\,0639-32 for a region around the Fe\,{\rm I} line at 543.452 nm.} 
\label{fig:bothDisentangledSpec}
\end{figure}

We have used our own implementation of the matrix disentangling algorithm of \cite{1994A&A...281..286S} to obtain individual spectra of the two stars. This method was adapted to include the apparent flux ratio between the stars for each observed spectrum as additional parameters in the disentangling. This allows the spectrum taken during the secondary eclipse to be included with a flux ratio of zero in the disentangling analysis. It is assumed that all spectra obtained out of eclipse have the same flux ratio. A simple grid search algorithm was used to optimise this parameter by finding the value of this flux ratio that minimises the root mean square (RMS) residuals between the observed spectra and the spectra reconstructed from the disentangled spectra. This search algorithm forms part of a Python wrapper that converts the UVES spectra into the appropriate format for disentangling algorithm. First, the wrapper re-interpolates the spectra onto a uniform logarithmic wavelength scale in the barycentric reference frame. A median filter is used as an initial normalisation, points affected by cosmic rays are replaced with a median value, and the spectrum is binned to provide the a sampling similar to the original UVES spectra. The disentangling is focused on small sections of spectrum, 30\,{\AA} in length, around specific Fe{\rm \,I} and Fe{\rm \,II} lines. By choosing a small region around each line, uncertainties from the continuum placement are reduced, compared with a fitting a continuum to the entire spectrum. The processing is done for each observed spectrum, and is done independently for each small wavelength section. Sections of spectrum were processed around each iron line present in the line-list of \cite{2013MNRAS.428.3164D}. Residuals between the observed spectra and reconstructed spectra were used to search for and remove remaining cosmic rays using three times the standard deviation of the residuals as a rejection criterion. The affected regions are replaced with the fitted values. The spectra were normalised and the disentangling algorithm is run one final time, using the optimal flux ratio. Figure \ref{fig:bothDisentangledSpec} shows an example of the resulting disentangled spectra around the Fe\,{\rm I} at 543.452\,nm, where $L_{2}/L_{1}=0.0513$.

\begin{figure*}
\centering
\resizebox{\hsize}{!}{\includegraphics[width=18cm,height=7cm]{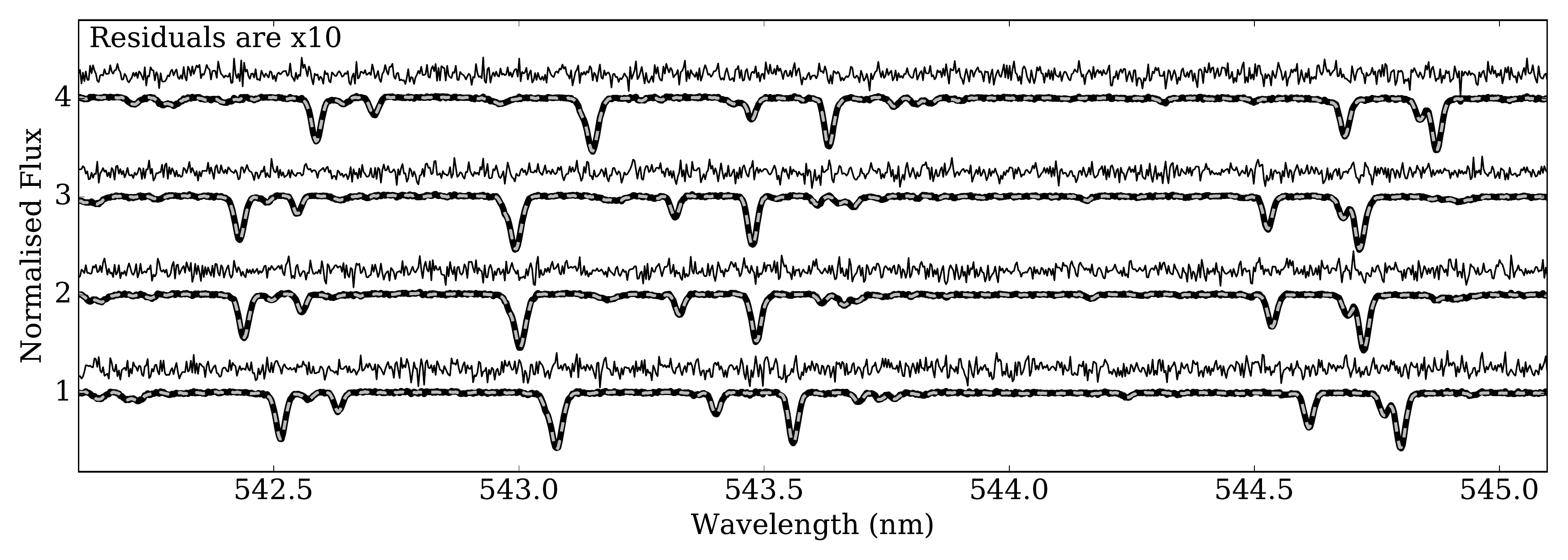}}
\caption{Observed (black) and recombined disentangled spectra (grey dashed) around the Fe\,{\rm I} line at 543.452 nm for four spectra from WASP\,0639-32. The residuals, scaled by a factor of 10, are shown 0.25 above each spectrum segment. Plotted with an offset of 1.0 between spectra.} 
\label{fig:disentangleResidual}
\end{figure*}

Figure \ref{fig:disentangleResidual} compares the observed and recombined disentangled spectra for four of the eleven observed spectra, around the same Fe\,{\rm I} line that is shown in Figure \ref{fig:bothDisentangledSpec}. The residuals, which have been scaled up by a factor of ten so they can be seen, are shown above the spectra. There are no strong features in the residuals. Large spikes near the absorption lines would indicate the radial velocities are incorrect, or large slopes at the end of a segment in the continuum of Figure \ref{fig:bothDisentangledSpec} would suggest there was an issue with the normalisation. The four spectra shown in Figure \ref{fig:disentangleResidual} are a representative sample of the set.

\subsection{Equivalent width fitting}
\label{subsec:EWfit}
 
An automated method is used to obtained a table of equivalent widths (EWs) for selected Fe line of each star, independently. 
Functions from the 2014 version of iSpec \citep{2014A&A...569A.111B} are used to first normalise and fit a continuum to each segment of disentangled spectrum. A small region around the expected location of the iron line is searched to identify the line, and cross-match the atomic data with the information in the \cite{2013MNRAS.428.3164D} line-list. In some cases multiple matches were found, in which case the one closest the expected wavelength was chosen. Once a line has been identified, a Gaussian profile is fitted to the line and is used to determine the equivalent width (EW) for a particular line. Each fit has been visually checked to ensure a sensible fit was made. Any lines that were blended or were significantly offset from the continuum were removed from the final selection equivalent widths. In total, there are 38 EWs for the primary (28 Fe\,{\rm I} lines and 10 Fe\,{\rm II} lines) and 19 EWs for the secondary star (16 Fe\,{\rm I} lines and 3 Fe\,{\rm II} lines). 

An EW fitting procedure, wrapped in the MCMC Python module, {\texttt{emcee}} \citep{2013PASP..125..306F}, was used to determine the best effective temperature, surface gravity, metallicity and microturbulence, ($T_{\rm eff}$, $\log g$, [Fe/H] and $v_{\rm mic}$, respectively) for each star separately. Initial walker positions are generated by perturbing initial parameter estimates by multiplying numbers drawn randomly from a normal distribution by step parameters (100, 0.05, 0.1 and 0.05 for $T_{\rm eff}$, $\log g$, [Fe/H] and $v_{\rm mic}$, respectively). The step parameters are needed because of the different scales associated with the different parameters. For a given set of parameters, \mbox{{\bf $\Theta$}$_{\rm sp}=\{T_{\rm eff}, \log g, {\rm [Fe/H]}, v_{\rm mic}\}$}, small sections of synthetic spectrum are generated around the lines used for the EW measurements using solar abundances from \cite{2009ARA&A..47..481A}, MARCS.GES \citep{MARCSModelAtmo} model atmospheres and the line-list and atomic data of \cite{2013MNRAS.428.3164D}. The projected rotational velocity, $v \sin i$, and macroturbulence, $v_{mac}$ are not fitted, and the iSpec's default parameters are used ($v \sin i=2$\,km\,s$^{-1}$ and $v_{\rm mac}=$3\,km\,s$^{-1}$) when generating the sections of synthetic spectrum. These parameters are not consider in the MCMC fitting because they do not alter the measured EW \citep{2014dap..book.....N}. Once measured, the EWs from the synthetic spectrum sections $W_{\rm s}$ are compared to the observed EWs, $W_{\rm o}$, to judge the choice of {\bf $\Theta$}$_{\rm sp}$. The overall log-likelihood function can be written as
\begin{equation}
\label{eq:ewloglike}
\ln \mathcal{L}({\bf W};{\bf \Theta}_{\rm sp}) = -\frac{1}{2}
\left[ \sum^N_{n=1} \left(\frac{W_{{\rm o},n}-W_{{\rm s,}n}({\bf \Theta}_{\rm sp})}{\sigma_{{\rm W,} n}^{2}} \right)^2 
- \ln \left( \frac{2\pi}{\sigma_{{\rm W,} n}^2}\right) \right]
\end{equation}
where {\bf{W}} is a vector of length $N$ containing the fitted equivalent widths, and $\sigma_{\rm W}$ are the uncertainties associated with the measured EWs. Priors were used to prevent the walkers exploring outside of the limits set by the model atmospheres, i.e. \mbox{$4500<T_{\rm eff}<7500$}, \mbox{$3.5< \log g < 5.0$}, \mbox{$-1.0< {\rm [Fe/H]}<1.0$} and \mbox{$0< v_{\rm mic}<100$}. For both the primary and secondary star, 100 walkers and 500 steps were used. For both stars, the first 100 steps were removed to allow for adequate burn-in. Auto-correlation lengths are between 50-70 for the different runs, and acceptance fractions are 0.34-0.55.

\begin{table*} 
\caption{Spectroscopic parameters for both components of WASP\,0639-32 obtained using equivalent width fitting, for cases where the surface gravity was free and where it was fixed at photometric values.}
\label{tab:j0639SpecParam}
\centering{
\begin{tabular}{l r r  r r}
\hline\hline
\noalign{\smallskip}
Parameter			& \multicolumn{2}{c}{Primary }		& \multicolumn{2}{c}{Secondary}  \\
				& \multicolumn{1}{c}{Free} 		& \multicolumn{1}{c}{Fixed} 			& \multicolumn{1}{c}{Free} 		& \multicolumn{1}{c}{Fixed}	\\
\noalign{\smallskip}
\hline
\noalign{\smallskip}
$T_{\rm eff}$ (K)			& $6730\pm30$	& $6320\pm10$	& $5490\pm100$	& $5420\pm90$	\\
$\log g_{\rm s}$				& $4.65\pm0.05$	& \multicolumn{1}{r}{$3.97$}			& $4.88\pm0.13$	& \multicolumn{1}{r}{4.61}	\\
$v_{\rm mic}$ (km s$^{\rm -1}$)& $1.61\pm0.01$	& $1.49\pm0.01$	& $2.67\pm0.18$	& $2.61\pm0.18$	\\
${\rm[Fe/H]}$				& $-0.10\pm0.01$	& $-0.33\pm0.01$	& $-0.38\pm 0.06$	& $-0.45\pm0.05$	\\
\noalign{\smallskip}
\hline
\end{tabular}}
\end{table*}

Table \ref{tab:j0639SpecParam} shows the resulting spectroscopic parameters for both stars in WASP\,0639-32, for two different cases. Parameter uncertainties are calculated from the 15.9, 50 and 84.1 percentiles. Where a single standard error is quoted, we have used the larger of the similar values for the positive and negative error bars. The first case, labelled ``free'', has the surface gravity included as a free parameter in the EW-fitting. In the second case, labelled ``fixed'', the surface gravity was fixed at the values shown in Table \ref{tab:MassRadiiData}. When included as a free parameter, the surface gravity of the primary is very different from the value obtained directly from the mass and radius, $\log g_{\rm MR} =3.974\pm 0.011$. Discrepancies between $\log g$ values derived from spectroscopic analysis (denoted $\log g_{\rm s}$ hereafter) and more direct independent techniques have been noted before. \cite{2013A&A...558A.106M} looked at 90 transiting planet host stars and showed that the discrepancy is temperature dependent. For a star with an effective temperature near 6700\,K, the difference is typically 0.6 dex, which is consistent with what we have found for the primary of WASP\,0639-32. 
In comparison, the difference between the two surface gravities for the secondary is smaller at 0.27\,dex. This is larger than the trend of \cite{2013A&A...558A.106M}, but maybe caused by the small number of Fe\,{\rm II} lines. \cite{2015PhDT........16D} notes that temperatures determined via the ionisation balance method are not affected by the discrepancy in $\log g$ but as our equivalent width fitting includes microturbulence and metallicity, the different surface gravities may affect these parameters. We therefore ran the equivalent width fitting procedure with $\log g_{\rm s}$ fixed at 3.97 and 4.61, for the primary and secondary, respectively. The results are also shown in Table \ref{tab:j0639SpecParam}. By fixing the surface gravities of both stars, the effective temperature of primary has been reduced, while it has increased for the secondary. For both stars, the metallicity has been reduced. The temperature ratio, $T_{2}/T_{1}$, has increased from $0.816\pm0.019$ to $0.857\pm0.016$. Note, the uncertainties on the values in Table \ref{tab:j0639SpecParam} do not include uncertainties from systematics such as those from using one specific line-list, solar abundance list or set of model atmospheres. \cite{2016arXiv161205013J} provide a summary of expected uncertainty contributions to the metallicity for systematic such as these, with one on the largest contributions being the continuum placement with a systematic of 0.3 dex. As the uncertainty on the effective temperature of the primary star is very small and does not include potential systematics (as previously discussed), we take the uncertainty as $\pm50$\,K. This has been estimated from variations in the temperature when the segment width for the generated spectrum is altered slightly. This paper has used the same line-list for both measuring the EW from the spectra and for generating the synthetic EWs. The same is true for the model atmospheres and solar abundances.

\begin{figure}
\centering
\resizebox{\hsize}{!}{\includegraphics{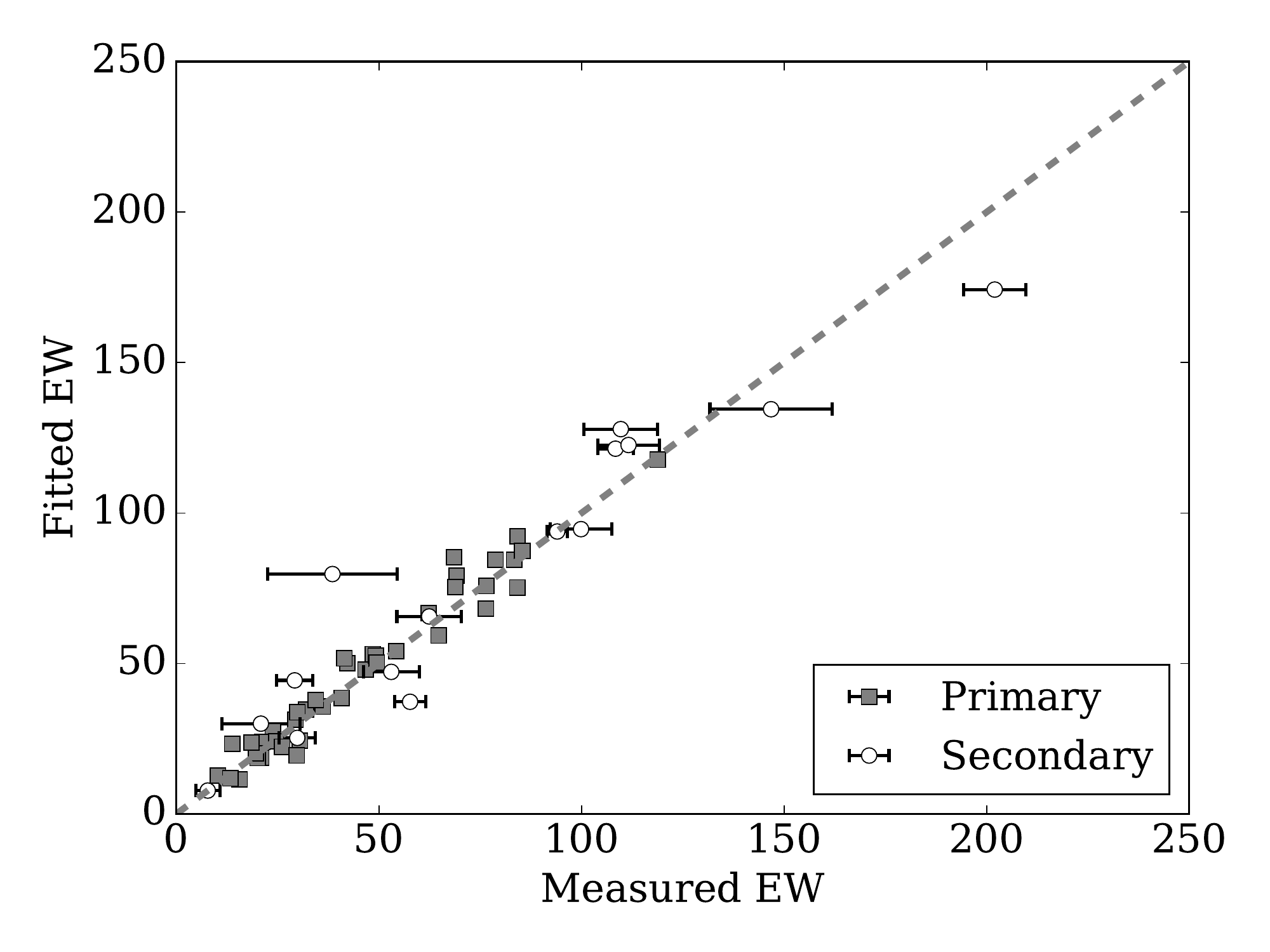}}
\caption{Comparison between fitted equivalent widths generated using the best-fitting spectroscopic parameters for a fixed $\log g$ (detailed in Table \ref{tab:j0639SpecParam}) and the measured equivalent widths for both components in WASP\,0639-32. Uncertainties on the primary equivalent widths are too small to see.} 
\label{fig:ewDiff}
\end{figure}

Figure \ref{fig:ewDiff} shows how the fitted EWs, generated with the parameters in Table \ref{tab:j0639SpecParam} where $\log g$ was fixed, compare to the measured values. Overall, there is good agreement with the measured values. Any lines that were significantly affected by telluric absorption were removed before the fitting, as were lines that were blended with other nearby lines.

To investigate how a poor continuum placement could or the disentangling process could affect the measured EWs, we have compared the EWs from a synthetic spectra generated with spectroscopic parameters similar to the two components of WASP\,0639-32, to the spectra obtained through disentangling. On average the disentangling altered the EWs of the primary star by less than 0.5\%, and 5\% for the secondary. This has not been included in the MCMC.

\subsection{Testing the equivalent width fitting method}
\label{subsec:testingEWmethod}

\begin{table} 
\caption{Recovered stellar parameters using equivalent width fitting, for equivalent widths measured from a synthetic spectrum.}
\label{tab:SyntheticParam}
\centering{
\begin{tabular}{l r r}
\hline\hline
\noalign{\smallskip}
Parameter			& \multicolumn{1}{c}{Recovered}		& \multicolumn{1}{c}{Target}  \\
\noalign{\smallskip}
\hline
\noalign{\smallskip}
$T_{\rm eff}$ (K)			& $6210\pm60$	& $6200$	\\
$\log g$				& $4.01\pm0.14$	& $4.0$	\\
$v_{\rm mic}$ (km s$^{\rm -1}$)& $1.53\pm0.09$	& $1.5$	\\
${\rm[Fe/H]}$				& $-0.05\pm0.03$	& $0.0$	\\
\noalign{\smallskip}
\hline
\end{tabular}}
{\tablefoot{Parameters used to generate the synthetic spectrum are labelled `Target'.}}
\end{table}

To ensure the equivalent width fitting was correctly recovering equivalent widths, we measured a set of EWs from a synthetically generated spectrum, and required that the estimated stellar parameters matched those used to generate the synthetic spectrum in iSpec. The target parameters used to generate the synthetic spectrum are listed in Table \ref{tab:SyntheticParam}. In addition to these parameters, we used The Vienna Atomic Line Database \citep[VALD,][]{2011BaltA..20..503K} line-list and atomic data, a resolution of $R=57\,000$ and a wavelength step of 0.001 over the range 500-700nm. A selection of 47 iron lines (42 Fe\,{\rm I} lines and 5 Fe\,{\rm II} lines) were chosen and EWs measured using the line fitting function within iSpec. Each line was checked to ensure there was no obvious blending. As these EWs were measured from a synthetic spectrum, an uncertainty is not automatically calculated. The EW-fitting method needs some uncertainties for the MCMC walkers are to explore parameter-space properly. As such, an uncertainty on 1.0\,m{\AA} was assumed for each EW. Using solar abundances from \cite{2009ARA&A..47..481A} and MARCS.GES \citep{MARCSModelAtmo} model atmospheres, these measured synthetic EW were fitted, and the resulting stellar parameters are shown in Table \ref{tab:SyntheticParam}. The MCMC used 100 walkers and 500 steps. Overall the recovered parameters agree well with those used to generate the synthetic spectrum. The metallicity is slightly lowered that expected but only just outside the 1-$\sigma$. 

We also wanted to ensure the EW fitting method could correctly recover stellar parameters of stars. We measured EWs from the ESPaDOnS spectrum of \object{Procyon} A in the Gaia Benchmark Star library \citep{2014A&A...566A..98B} and used the EW fitting method described in Sect. \ref{subsec:EWfit} to estimate its stellar parameters. The spectrum was first normalised, by dividing through with the continuum. The spectrum was split into segments and a third-degree polynomial was fitted in each segment in order to identify the continuum. The VALD line-list \citep{2011BaltA..20..503K} was used to select 87 lines (72 Fe\,{\rm I} and 11 Fe\,{\rm II} lines) which, from visual inspection, did not appear to be blended. For the EW fitting, again we used solar abundances from \cite{2009ARA&A..47..481A} and MARCS.GES \citep{MARCSModelAtmo} model atmospheres. The resolution of the generated spectrum was adjusted to account for the slightly higher resolving power ($R=65000$ instead of $R=56990$). Here, the MCMC used 200 walkers with 1000 steps each. The first 300 were discarded as a burn-in stage. For this particular run we allowed both a Voigt and Gaussian profile to be used to determine the synthetic EW, and chose the EW from the best fitting profile. As a result, the MCMC required additional steps but it allowed an improved match to the spectral lines. The results and Gaia benchmark parameters \citep{2015A&A...582A..49H,2014A&A...564A.133J} are shown in Table \ref{tab:ProcyonParam}. The parameters themselves are taken as the median value from the distribution, with uncertainties calculated using the 15.9 and 84.1 percentiles. Overall the measured effective temperature, surface gravity and metallicity agree with the literature values. The microturbulence is within 2-$\sigma$ of the literature value, which is itself a mean value obtained from multiple techniques.

\begin{table} 
\caption{Spectroscopic parameters for Procyon obtained using equivalent width fitting.}
\label{tab:ProcyonParam}
\centering{
\begin{tabular}{l r r}
\hline\hline
\noalign{\smallskip}
Parameter			& \multicolumn{1}{c}{This work}		& \multicolumn{1}{c}{Literature}  \\
\noalign{\smallskip}
\hline
\noalign{\smallskip}
$T_{\rm eff}$ (K)			& $6540\pm150$	& $6554\pm84$	\\
$\log g$					& $3.94\pm0.10$	& $4.0\pm0.02$	\\
$v_{\rm mic}$ (km s$^{\rm -1}$)& $1.52\pm0.16$	& $1.8\pm0.11$		\\
${\rm[Fe/H]}$				& $0.00\pm0.07$	& $+0.01\pm0.08$\parbox{0pt}{$^*$}	\\
\noalign{\smallskip}
\hline
\end{tabular}}
\tablefoot{\tablefoottext{*}{Uncertainty in [Fe/H] from combining different sources of uncertainties presented in Table 3 of the \cite{2014A&A...564A.133J} paper, in quadrature. This is the same approach as in \cite{2015A&A...582A..49H}.}}
\end{table}

\subsection{Checking effective temperatures}
\label{subsec:TeffChecks}

The equivalent width fitting gives two slightly different effective temperatures depending on whether we choose to use surface gravities from the spectroscopy, $\log g_{\rm s}$, or from the masses and radii, $\log g_{\rm MR}$. Therefore other methods of determining the effective temperature have been considered.

\begin{figure}
\centering
\resizebox{\hsize}{!}{\includegraphics{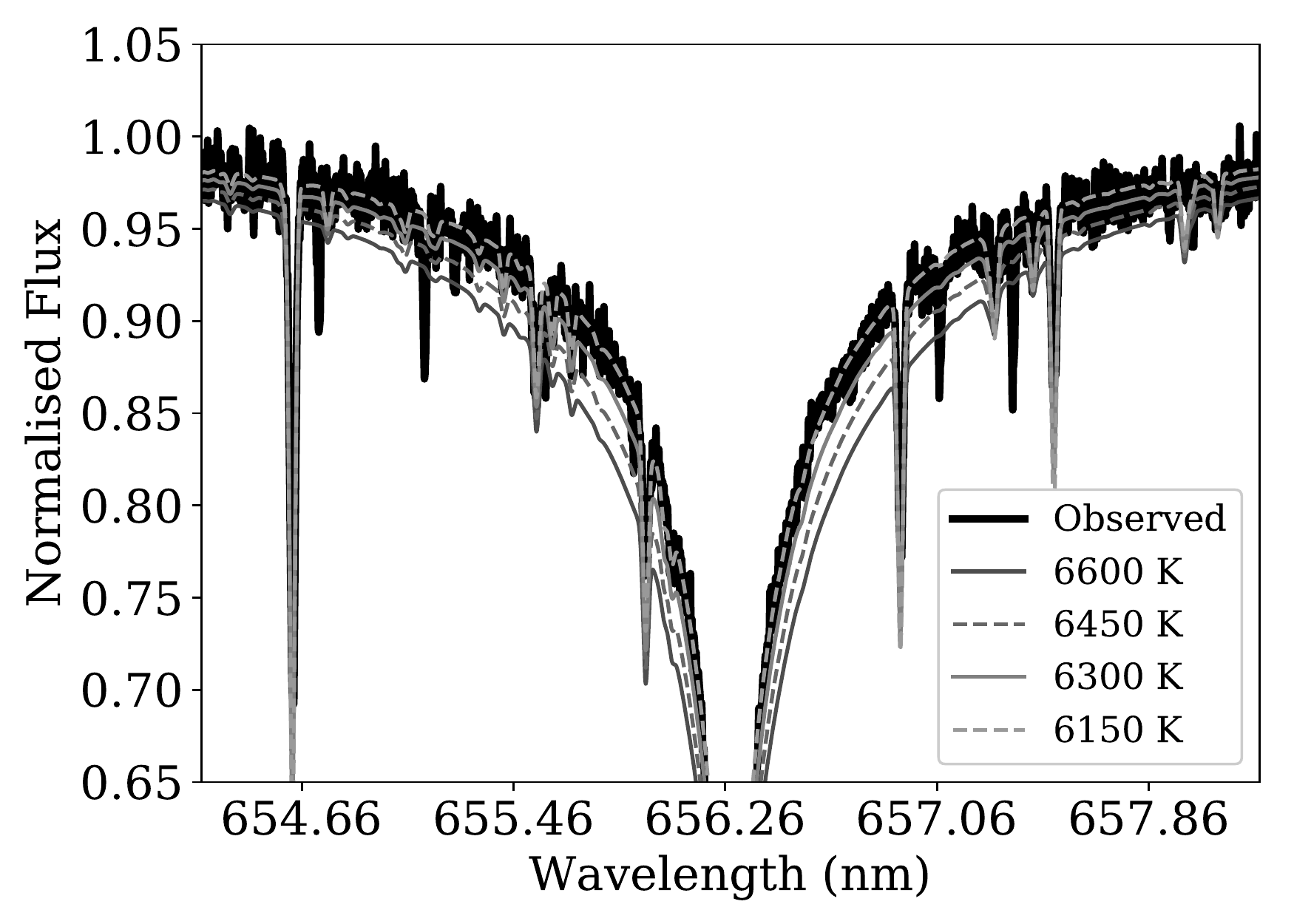}} 
\caption{Synthetic spectra of different temperatures plotted against the $H_{\alpha}$ wings of the primary star.} 
\label{fig:HalphaFit}
\end{figure}

One of the spectra for WASP\,0639-32, (taken at HJD = 2\,457\,042.59910) was taken during the flat region of the secondary eclipse when only the primary star would have been visible, therefore, it can treated in the same way as a single star spectrum. Figure \ref{fig:HalphaFit} shows how multiple synthetic spectra, generated using different effective temperatures, compare to the $H_{\alpha}$ region of the eclipse spectrum. By looking at the wings of the $H_{\alpha}$ line, we found the best temperature was $6150\pm150$\,K, which is lower than both estimates from the EW fitting. A surface gravity of $3.97$ was used to generate the synthetic spectra, but as this method is insensitive to the surface gravity \citep{2014dap..book.....N}, the chosen value will not affect the best model spectrum. [Fe/H] was set to $-0.15$ and microturbulence was set at 1.56\,km\,s$^{-1}$. The temperature obtained from the $H_{\alpha}$ line disagrees with both temperatures for the primary star in Table \ref{tab:j0639SpecParam}. Balmer line fitting can be very sensitive to continuum placement, something that can be difficult to do for \'{e}chelle spectra due to errors from the merging of different \'{e}chelle orders.  

We also fitted $H_{\alpha}$ lines from three different spectra of Procyon, that were also taken with the UVES instrument, to investigate if the offset could be caused by something in the UVES pipeline. The three spectra all had the same grating as spectral range and those for WASP\,0639-32, and they have been reduced using the same reduction pipeline. Two spectra were taken from the 092.D-0207(A) program from 2013, and one spectrum from the 266.D-5655(A) program from 2002. All three spectra produced temperatures that were consistent with the literature value shown in Table \ref{tab:ProcyonParam}, however in one case (one of the two spectra taken on the same night) the best temperature was 100\,K lower than the literature value. For this particular spectrum, the continuum fitting was more difficult due increased noise in the spectrum. As such uncertainties from the continuum placement (at least $\pm100$\,K) should be fed into the uncertainty on the temperature.

As a separate test, the EW-fitting was applied to lines from the spectrum taken in eclipse. Overall, 45 lines were chosen (40 Fe\,{\rm I} and 5 Fe\,{\rm II} lines), using the same model atmospheres and line list. As before, the MCMC used 100 walkers and 500 steps, with the initial 100 steps removed as burn-in, and the same checks were carried out to ensure suitable mixing and convergence. With the surface gravity fixed at 3.97, the resulting parameters were, $T_{\rm eff} = 6410\pm130$\,K, [Fe/H]$=-0.14\pm0.30$ and $v_{\rm mic}=1.42\pm0.80$\,km\,s$^{-1}$. While the uncertainties are much larger than those obtain through the disentangling, the effective temperature has better agreement with the temperature from the disentangling with fixed surface gravity rather than the temperature from the $H_{\alpha}$ fitting. 

\begin{table} 
\caption{Magnitudes used in the fitting with colour-effective temperature and colour-surface brightness relations.}
\label{tab:ObservedMag}
\centering{
\begin{tabular}{l r r}
\hline\hline
\noalign{\smallskip}
Band 			& \multicolumn{1}{c}{Magnitude}		& \multicolumn{1}{c}{Error}  \\
\noalign{\smallskip}
\hline
\noalign{\smallskip}
APASS {B}			& $11.193$	& $0.028$	\\
APASS {V} & 10.686 & 0.067 \\
APASS {g'} & 10.907 & 0.037 \\
APASS {r'} & 10.647 & 0.021\\
APASS {i'} & 10.437 & 0.050\\ 
TYCHO {B$_{\rm T}$} & 11.265 & 0.057 \\
TYCHO {V$_{\rm T}$} & 10.718 & 0.054 \\ 
2MASS {J} & 9.591 & 0.023 \\
2MASS {H} & 9.321 & 0.026 \\ 
2MASS {K$_{\rm s}$} & 9.306 & 0.019 \\
DENIS {I} & 10.128 &0.04	\\
DENIS {J} & 9.578 & 0.05 \\
DENIS {K} & 9.251 & 0.09 \\
WISE {W$_{\rm 3}$} & 9.222 & 0.031\\
\noalign{\smallskip}
\hline
\end{tabular}}
\end{table}

We have also used empirical colour--effective temperature and colour--surface brightness relations to estimate the effective temperatures of the individual stars in the binary system, as described in \cite{2018arXiv180310522M}. Photometry was taken from the AAVSO Photometric All-Sky Survey, APASS (B,V, g',r' and i'), the Two Micron All Sky Survey, 2MASS (JHK$_{\rm s}$), Tycho-2 Catalogue (B$_{\rm T}$ and V$_{\rm T}$), Deep Near-infrared Southern Sky Survey, DENIS, (I, J, K) and the Wide-Field Infrared Survey Explorer, WISE (W$_{3}$).  See Table \ref{tab:ObservedMag} for the observed values \citep{2009AAS...21440702H, 2006AJ....131.1163S, 2000A&A...355L..27H, 1997Msngr..87...27E, 2010AJ....140.1868W}.

To fit the observed photometry, the model uses the following parameters-- g$^{\prime}_{0, i}$, the apparent g$^{\prime}$-band magnitudes for stars $i=1$, $i=2$ corrected for extinction; $T_{\rm eff,i}$ the effective temperatures for each star; E$({\rm B}-{\rm V})$, the reddening to the system; $\sigma_{\rm ext}$ the additional systematic error added in quadrature to each measurement to account for systematic errors. Although not used here, the model can work with triple systems, in which case there is a third g$^{\prime}$-band and $T_{\rm eff}$. The apparent magnitudes for each star are predicted through the colour-temperature relations of \citet{2013ApJ...771...40B}. Following the example of  \citet{2013ApJ...771...40B}, the transformations of \cite{1988PASP..100.1134B} and \cite{2001AJ....121.2851C} are used to transform between the photometric systems of Johnson and 2MASS. Cousins I$_{\rm C}$ is used as an approximation to the DENIS Gunn i$^{\prime}$ band and the 2MASS K$_{\rm s}$ is used as an approximation to the DENIS {K} band. Transformations between Johnson {B},{V} and Tycho B$_{\rm T}$ and V$_{\rm T}$ magnitudes are carried out by interpolating values in Table 3 of \cite{2000PASP..112..961B}. {V}-band extinction is assumed to be $3.1\times {\rm E}({\rm B}-{\rm V})$, extinction in the SDSS and 2MASS bands is calculated using A$_{\rm r} = 2.770\times {\rm E}({\rm B}-{\rm V})$ from \cite{2003A&A...401..781F}, and the extinction coefficients to the r$^{\prime}$ band come from \citet{2014MNRAS.440.3430D}. The average surface brightness ratio in the WASP band ($0.429\pm0.007$) as an approximation to the {V}-band ratio, and allowed the external error $\sigma_{\rm ext}$ account for differences in the two bands. This ratio is used as a constraint, by using the {V}-band surface brightness and {{B-K}} relations of \cite{2017ApJ...837....7G}. The parameter-space was explored using a maximum likelihood approach, using an {\texttt{emcee}}, \citep{2013PASP..125..306F} MCMC procedure,  with 64 walkers, 1088 steps and 64 burn-in steps. All chains have visually inspected to ensure suitable mixing, and via a running mean check to ensure convergence. The median acceptance fraction for the walkers was 0.392. The total line-of-sight reddening from the dust maps of \cite{2011ApJ...737..103S} is used to place a prior on $\Delta = {\rm E}({\rm B}-{\rm V}) - {\rm E}({\rm B}-{\rm V})_{\rm map}$:
 \[ P(\Delta) =\left\{ \begin{array}{ll} 1 & 
\Delta \le 0 \\ \exp(-0.5(\Delta/0.034)^2) &
\Delta > 0 \\ \end{array} \right. \]
The 0.034 is a constant from \cite{2014MNRAS.437.1681M}, and is the offset between reddening maps and the values derived from photometry of a number of A-type stars.

\begin{figure}
\centering
\resizebox{\hsize}{!}{\includegraphics{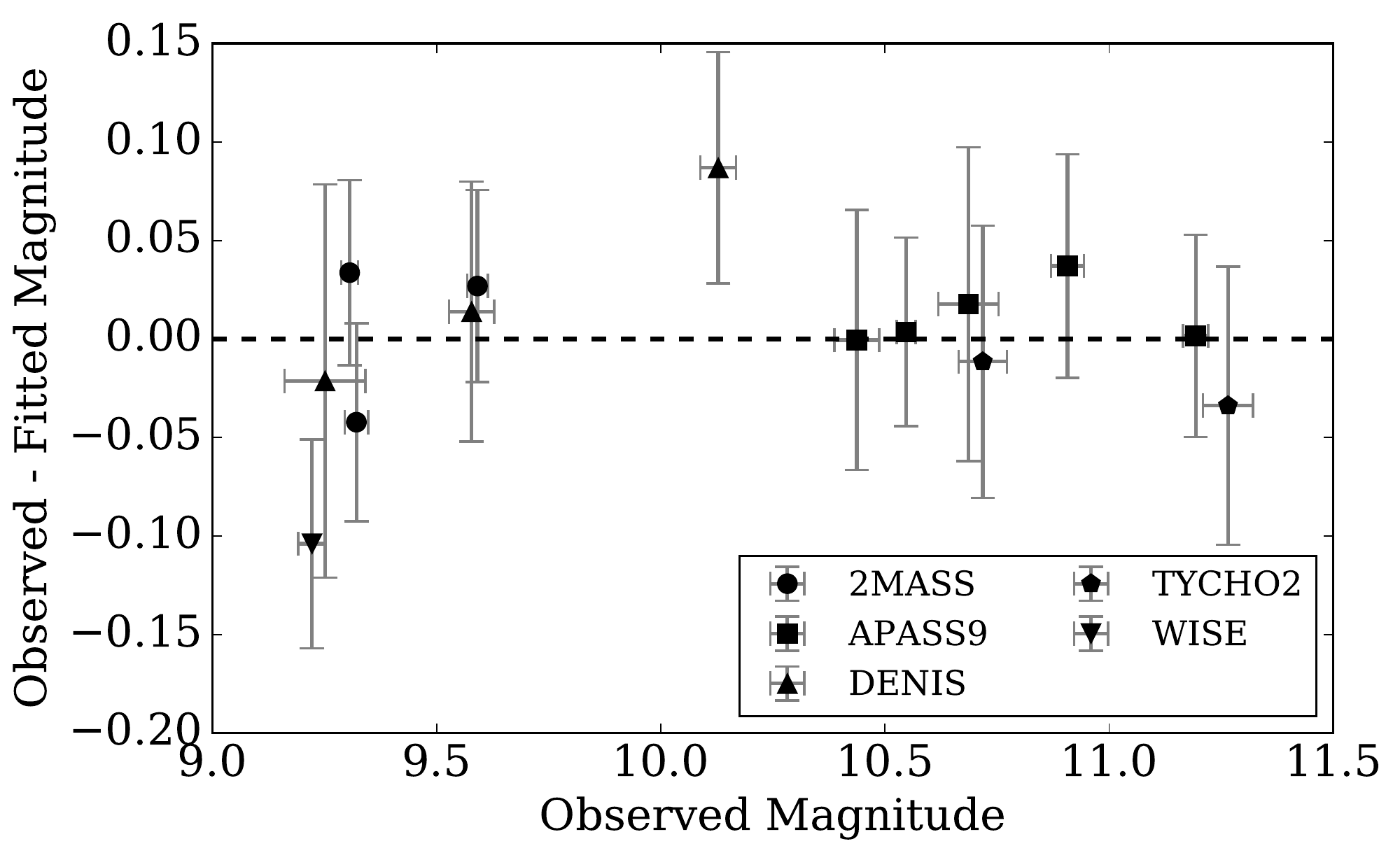}}
\caption{Difference between the observed magnitude and those obtained through the fitting of colour-temperature relations. Uncertainties on the y-axis include the external error, $\sigma_{\rm ext}$, added in quadrature.}
\label{fig:magFit}
\end{figure}

	The resulting temperatures were $T_{1}=6340\pm{190}$\,K and $T_{2}=5320\pm{170}$\,K with reddening of ${\rm E}({\rm B}-{\rm V})=0.067\pm{0.036}$ and $\sigma_{\rm ext} = 0.043\pm0.017$. The parameter values are as the median of the posterior distributions, while the uncertainties are the standard deviations. Figure \ref{fig:magFit} shows the residuals from the fit.
	
Overall, by taking the weighted mean of the temperatures from each technique for each star (excluding spectroscopy with a free surface gravity), the effective temperatures are found to be $6330\pm 50$\,K and $5400\pm80$\,K for the primary and secondary, respectively. Note that in these calculations, an additional 100K has been added to the uncertainty in the temperature for the primary star obtained from the $H_{\alpha}$ fitting, due to the uncertainty in the continuum placement, and the uncertainty from EW fit with a fixed $\log\,g$ has been taken as 50\,K.

\section{Absolute parameters}
\label{sec:massAndRadii}

\begin{table} 
\caption{Absolute parameters for WASP\,0639-32.}
\label{tab:MassRadiiData}
\centering{
\begin{tabular}{l r r}
\hline\hline
\noalign{\smallskip}
Parameter				& Primary & Secondary \\
\noalign{\smallskip}
\hline
\noalign{\smallskip}
Period (days)				& \multicolumn{2}{c}{$11.658317(5)$}	\\
$e$						& \multicolumn{2}{c}{$0.0009(^{+12}_{-06})$}			\\
$i$ (\degr)					& \multicolumn{2}{c}{$89.9943(67)$}		\\
$a \sin i$ (R$_{\sun}$)		& \multicolumn{2}{c}{$26.964(31)$}	\\
$a$ (R$_{\sun}$)		& \multicolumn{2}{c}{$26.964(31)$}	\\
Mass ratio					& \multicolumn{2}{c}{$0.6785(16)$}  	 \\
$M \sin^{3}i$ (M$_{\sun}$)	& $1.1544(43)$ 	& $0.7833(28)$	 		   \\
Mass (M$_{\odot}$)			& $1.1544(43)$		& $0.7833(28)$		\\
Radius (R$_{\odot}$)			& $1.834(23)$		& $0.7291(81)$		  \\
$\log g_{\rm MR}$			& $3.974(11)$		& $4.607(10)$	\\ 
$T_{\rm eff}$ (K)			& $6330(50)$		& $5400(80)$	\\
$\log(L/L_{\odot})$			& $0.685(18)$		& $-0.392(28)$	\\
$d_{\rm K}$ (pc)			& \multicolumn{2}{c}{$323(06)$}		\\
Parallax (mas)				& \multicolumn{2}{c}{$3.10(23)(30)$}	\\
$d_{\rm gaia}$ (pc)			& \multicolumn{2}{c}{$323(^{+26}_{-22})(^{+35}_{-29}) $}		\\
$v \sin i$ \mbox{(km s$^{-1})$}	& $6.75(7)$ 		& $4.7(1.1)$	\\
$v_{\rm synch}$ \mbox{(km s$^{-1})$} & $7.96(10)$		& $3.16(4)$	\\
\noalign{\smallskip}
\hline
\end{tabular}}
{\tablefoot{Where two uncertainties are given, the first refers to the random error, and the second is the systematic.}}
\end{table}

{\textsc{jktabsdim}}\footnote{\tt{http://www.astro.keele.ac.uk/jkt/codes/jktabsdim. html}} was used to calculate the absolute parameters for WASP\,0639-32, taking the relative radii and inclination from the detrended fit in Table \ref{tab:DetrendData}, and the semi-amplitude velocities and eccentricity from Table \ref{tab:j0639orbitParam}. The results are presented in Table \ref{tab:MassRadiiData}. $\log g_{\rm MR}$ is used to denote a surface gravity obtained directly from the mass and radius of the star. {\textsc{jktabsdim}} also calculates a distance estimate using the surface-brightness relations of \cite{2004A&A...426..297K}. The 2MASS {\em K$_{\rm s}$} apparent magnitude has been used to estimate a distance, $d_{\rm K}$, to WASP\,0639-32 and is shown in Table \ref{tab:MassRadiiData}. The magnitude was converted to the Johnson system using equations from \cite{1988PASP..100.1134B} and \cite{2001AJ....121.2851C}, and the APASS {\em V}-band magnitude, giving $K=9.267\pm0.097$ mag. Although other apparent magnitudes are available and give similar distances, for example, $d_{\rm B}=310\pm23$\,pc and $d_{\rm V}=322\pm21$\,pc for APASS {\em B,V} respectively, the {\em {K}}-band distance was chosen as it is the least sensitive to reddening. Also listed in this table, is the parallax from Gaia DR1 \citep{2016A&A...595A...2G} and its associated Gaia distance, $d_{\rm gaia}$ for comparison. The distance from the $K$-band magnitude agrees well with $d_{\rm gaia}$.

\begin{figure}
\centering
\resizebox{\hsize}{!}{\includegraphics{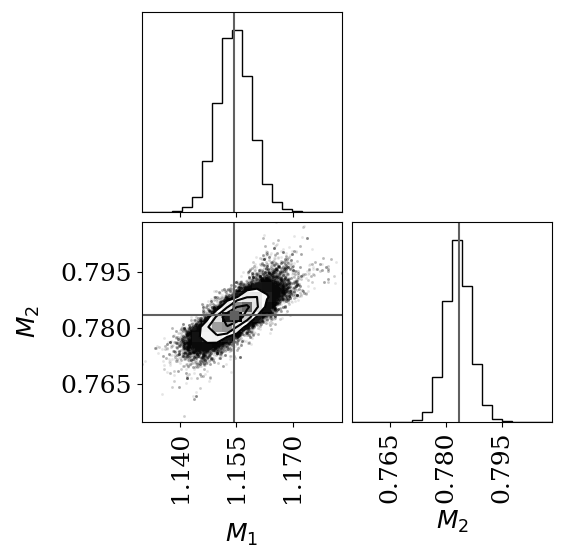}}
\caption{Density distribution showing the correlation between $M_{1}$ and $M_{2}$. A grey cross marks the measured values as shown in Table \ref{tab:MassRadiiData}.}
\label{fig:massCorrelation}
\end{figure}

The quoted uncertainties in the masses are 0.37\% and 0.36\% for $M_{1}$ and $M_{2}$, respectively, while the uncertainties in the radii are 1.27\% and 1.12\% for $R_{1}$ and $R_{2}$, respectively. For the masses, this is the same level of precision as was obtained for the evolved binary, AI Phe \citep{2016A&A...591A.124K}, meaning that like AI Phe, the parameters of WASP\,0639-32 will be suitable for constraining the helium abundance in stellar evolutionary models. Figure \ref{fig:massCorrelation}\footnote{Produced using {\texttt{corner.py} \citep{corner}.}} shows the correlation between $M_{1}$ and $M_{2}$, calculated from $K_{1}$, $K_{2}$ and $e$ values explored during the MCMC in Sect. \ref{sec:RVAnalysis}. The formulae recommended in \cite{2010A&ARv..18...67T} were used for the calculation as they are same as the formulae used in {\textsc{jktabsdim}}. The period and inclination were fixed for this calculation.

The projected rotational velocity, $v \sin i$, of both stars has been measured using the cross-correlation technique \citep{2013EAS....62..143B}. The spectroscopic parameters from the fixed $\log g$ solution in Table \ref{tab:j0639SpecParam}, were used to generate synthetic spectra with $v \sin i $ values ranging from 2.0 to 9.0 km\,s$^{-1}$. The spectra were generated using solar abundances from \cite{2009ARA&A..47..481A}, MARCS.GES \citep{MARCSModelAtmo} model atmospheres and the VALD line-list \citep{2011BaltA..20..503K}. Each synthetic spectrum was cross-correlated with the NARVAL solar spectrum \citep{2003EAS.....9..105A} within iSpec to set up a relation between $v \sin i$ and the full-width half-maximum (FWHM) of the cross-correlation peak. Each observed spectrum was cross-correlated with the same solar spectrum and the FWHM was obtained for both stars. After taking the mean and standard deviation of the resulting FWHMs, the established relation for each star was used to find the corresponding $v \sin i$, the results of which are shown in Table \ref{tab:MassRadiiData}. Also included in Table \ref{tab:MassRadiiData} are the synchronous velocities $v_{\rm synch}$ for both stars, which are calculated in {\textsc{jktabsdim}}. Comparisons between $v \sin i$ and $v_{\rm synch}$ show that both stars are almost rotating synchronously. For the primary star the difference is sufficient to say that is sub-synchronous, although there maybe systematic uncertainties that have not been taken into consideration, which may alter the outcome. One example would be the effect of microturbulence in the synthetic spectra. A possible explanation for the nature of the primary is that it is evolving on a timescale that is more rapid than the timescale for synchronisation.

\section{Implication for stellar evolutionary models}
\label{sec:models}

\subsection{The models}
\label{subsec:modsDescripts}
With uncertainties on the masses of less than 0.5\% and one component evolved off of the main-sequence, WASP\,0639-32 is a good binary system for testing stellar evolutionary models \citep{2010A&ARv..18...67T}. The model tracks that we tested, were produced using the Garching Stellar Evolution Code, \citep[{\textsc{garstec}},][]{2008Ap&SS.316...99W} and include the initial helium abundance, $Y_{\rm i}$, initial metallicity, $Z_{\rm i}$, and mixing-length parameter, $\alpha_{\rm ml}$, as free parameters.

The model grid covers six initial helium abundances, $Y_{\rm i}=\{0.231, 0.251, 0.271, 0.291, 0.311, 0.331\}$ and five different mixing length parameters, $\alpha_{\rm ml} = \{1.598,1.698,1.798,1.898,1.989 \}$, yielding 30 combinations in total. For each pair of $Y_{\rm i}$ and $\alpha_{\rm ml}$, there are 15 values for $Z_{\rm i}$, covering a range from 0.00307 to 0.0772 with equal spacing in $\log Z$ space. For each point on an evolutionary track, [Fe/H]$_{\rm s}$ is calculated using [Fe/H]$_{\rm s} = \log_{10}(Z_{\rm s}/X_{\rm s})-\log_{10}(0.02439)$, where 0.02439 is the solar value from \cite{1993oee..conf...15G}. $Z_{\rm s}$ and $X_{\rm s}$ are the model metal fraction and model hydrogen fraction, respectively. The evolutionary tracks cover a mass range of 0.7-2.0\,M$_{\sun}$ in steps of 0.02\,M$_{\sun}$.

The microphysics used in these models is described in \cite{2013MNRAS.429.3645S} and \cite{2008Ap&SS.316...99W}, but we provide a brief summary here. The convection is described by the standard mixing length theory of \cite{1990sse..book.....K}, where the solar mixing length is $\alpha_{\rm ml,\odot} = 1.801$ using the \cite{1993oee..conf...15G} solar composition. Due to the effects of diffusion, the initial solar composition is found to be ${\rm [Fe/H]}_{\rm i}= +0.06$. Molecular opacities from \cite{2005ApJ...623..585F} complement the OPAL opacities from \cite{1996ApJ...464..943I}. Convective mixing is described in terms of a diffusive process, with the diffusion coefficient, $D_{\rm c}$, being given by
\begin{equation}
\label{eq:diffCoeff}
D_{\rm c} = \frac{1}{3} \alpha_{\rm ml}H_{\rm p} v_{c}
\end{equation}
where $v_{\rm c}$ is the convective velocity and $\alpha_{\rm ml} H_{\rm p}$ is the local mixing length. Convective overshooting is described by the diffusive process \citep{1996A&A...313..497F} given by 
\begin{equation}
\label{eq:diffConvec}
D = D_{0} \exp \left( \frac{-2z}{f h_{\rm p}} \right)
\end{equation}
where $f$ is a free parameter dictating the extent of the overshooting, $D_{0}$ is the diffusion coefficient within the convective border and $z$ is the distance to the convective border. $h_{\rm p}$ is a function of the pressure scale height ($H_{\rm p}$) and the thickness of the convective core ($\Delta R_{\rm CZ}$) given as 
\begin{equation}
\label{eq:hp}
h_{\rm p} = H_{\rm P} \times {\rm min} \left[1, \left( \frac{\Delta R_{\rm CZ}}{H_{\rm P}} \right)^{2}  \right]
\end{equation}
This definition of $h_{\rm p}$ ensures the overshooting region is limited geometrically to a portion of the convective region, in cases where the convective region is small e.g. where $\Delta R_{\rm CZ}< H_{\rm p}$ \citep{2010ApJ...718.1378M}. If $\Delta R_{\rm CZ}> H_{\rm p}$, the geometric limit does not have an effect. In this case, our choice of $f=0.020$ leads to an extension of the overshoot $\sim0.25H_{\rm p}$.

The description of atomic diffusion is the same as is described in \cite{2016A&A...591A.124K}, and follows the equations of \cite{1969fecg.book.....B} and the method of \cite{1994ApJ...421..828T}. The parameterisation of \cite{2012ApJ...755...15V} is used to include extra macroscopic mixing by extending the convective envelope. For main-sequence stars, the effects from stellar winds and radiative acceleration are not considered in the models. These effects become relevant at higher masses and work to limit the atomic diffusion. Without the mitigating effects of extra mixing or winds, metals on the stellar surface, would become depleted by an amount that is greater than what is observed \citep{2014A&A...562A.102O}.

\subsection{Input data}
\label{subsec:input}

For each star in WASP\,0639-32, the vector of observed parameters is \mbox{\vec{d} = ($T$, $\rho$, $M$, [Fe/H]$_{\rm s}$)}, where $T$ is the effective temperature, $\rho$ is the mean density, $M$ is the mass and [Fe/H]$_{\rm s}$ is the observed surface metal abundance. Using 
\begin{equation}
\rho_{\rm n} = \frac{3 \pi}{GP^2(1+Q_{\rm n})}\left(\frac{a}{R_{\rm n}}\right)^{3} ,
\label{eq:stellarDensity}
\end{equation}
the mean density is a quantity that can be calculated from the fractional radii $r_{n}=(R_{\rm n}/a)$ obtained from the lightcurve solution and Kepler's laws, meaning it is independent from the mass estimates obtained from the spectroscopic orbit. In this equation, $R$ is the radius for star $n=1,2$, $a$ is the semi-major axis of the orbit, $P$ is the orbital period and $G$ is Newton's gravitational constant \citep{2015A&A...575A..36M}. $Q_{\rm n}$ is a function of the mass ratio ($q=M_{2}/M_{1}$), where $Q_{1} = q$ and $Q_{2} = 1/q$. The densities of the two stars were found to be \mbox{$\rho_{1}=0.1873\pm0.0071\,\rho_{\sun}$} and \mbox{$\rho_{2}=2.023\pm0.073\,\rho_{\sun}$} for the primary and secondary respectively. By using $M$ and $\rho$ rather than $M$ and $R$, the parameters in the log-likelihood can be assumed to be uncorrelated. The masses used are those stated in Table \ref{tab:MassRadiiData} and the temperatures are the weighted means that calculated at the end of Sec. \ref{subsec:TeffChecks}. We use the metallicities from Table \ref{tab:j0639SpecParam} using a fixed $\log g_{\rm s}$, but assume uncertainties of $\pm0.1$\,dex to account take into consideration some of the uncertainties mentioned in \citet{2016arXiv161205013J}. The grid of models described in Sect. \ref{subsec:modsDescripts} is quite coarse compared with the uncertainties on the masses and densities, and there is no interpolation between the models. As such, the uncertainties were increased to 0.0073 and 0.0050, for $M_{1}$ and $M_{2}$ respectively, to allow more evolutionary tracks to be explored. An uncertainty of 0.005\,$M_{\sun}$, covers a range of $\pm2$-$\sigma$ from the centre of a mass bin.

\subsection{Bayesian analysis}
\label{subsec:bayesian}

In order to explore this complex parameter space set out by the models, we use a Bayesian framework. The set of model parameters is given by \mbox{$\vec{m} = \left(\mbox{$\tau$}, \mbox{$M$}, \mbox{$\alpha_{\mathrm ml}$}, \mathrm{[Fe/H]}_{\mathrm{i}}\right)$}, where $\tau$, $M$, $\alpha_{\rm ml}$ and $\rm{[Fe/H]}_{\rm{i}}$ are the age, mass, mixing length and initial metal abundance respectively. The initial metal abundance differs from the observed surface abundance due to diffusion.

The probability distribution function  \mbox{$p(\vec{m}|\vec{d})$} is proportional to \mbox{${p(\vec{m})\cal L}(\vec{d}|\vec{m})$}. {$\cal L$}$(\vec{d}|\vec{m}) = \exp (-\chi^{2}/{2})$ is the likelihood of observing the data $\vec{d}$ given the model ${\vec{m}}$, assuming the independent observed parameters and Gaussian uncertainties, where 
\begin{eqnarray}
\label{eq:modelChisq}
\chi^2 =  \frac{\left(M - M_{\mathrm{obs}}\right)^2}{\sigma_{M}^2} 
+ \frac{\left(\rho - \rho_{\mathrm{obs}}\right)^2}{\sigma_{\rho}^2}
+ \frac{\left(T - T_{\mathrm{obs}}\right)^2}{\sigma_{T}^2} \\
 + \frac{\left(\mathrm{[Fe/H]}_{\mathrm{s}} - \mathrm{[Fe/H]}_{\mathrm{s,obs}}\right)^2}{\sigma_{\mathrm{[Fe/H]_{\mathrm{s}}}}^2} \nonumber.
\end{eqnarray}
Observed parameters are denoted by the ``obs'' subscript and each $\sigma$ represent the uncertainty of the parameter. $p(\vec{m})$ is the product of the priors on each of the model parameters given by $p(\vec{m}) = p(\tau)p(M)p($[Fe/H$]_{\rm i })$. A flat prior is applied to the initial surface metallicity, although this usually has little effect as the surface metallicity is constrained by the observed value. A very loose prior is used on the age aimed at keeping the star's age within the age of the universe while allowing room for the models to explore slightly older ages in order provide a more realistic age estimate. The prior on the age is set to $0< \tau<17.5$\,Gyr. 

Each $Y_{\rm i}$ value was considered independently and calculations were split into slices of age covering 300\,Myr, e.g. 0-300\,Myr, 300-600\,Myr, etc. and the age is the weighted average over the bin. This is to allow comparisons between the two stars and to choose results where age and initial helium abundance were common to both star.

\subsection{Model comparisons}
\label{subsec:ModelComparisons}

 \begin{figure*}
\centering
\resizebox{\hsize}{!}{\includegraphics{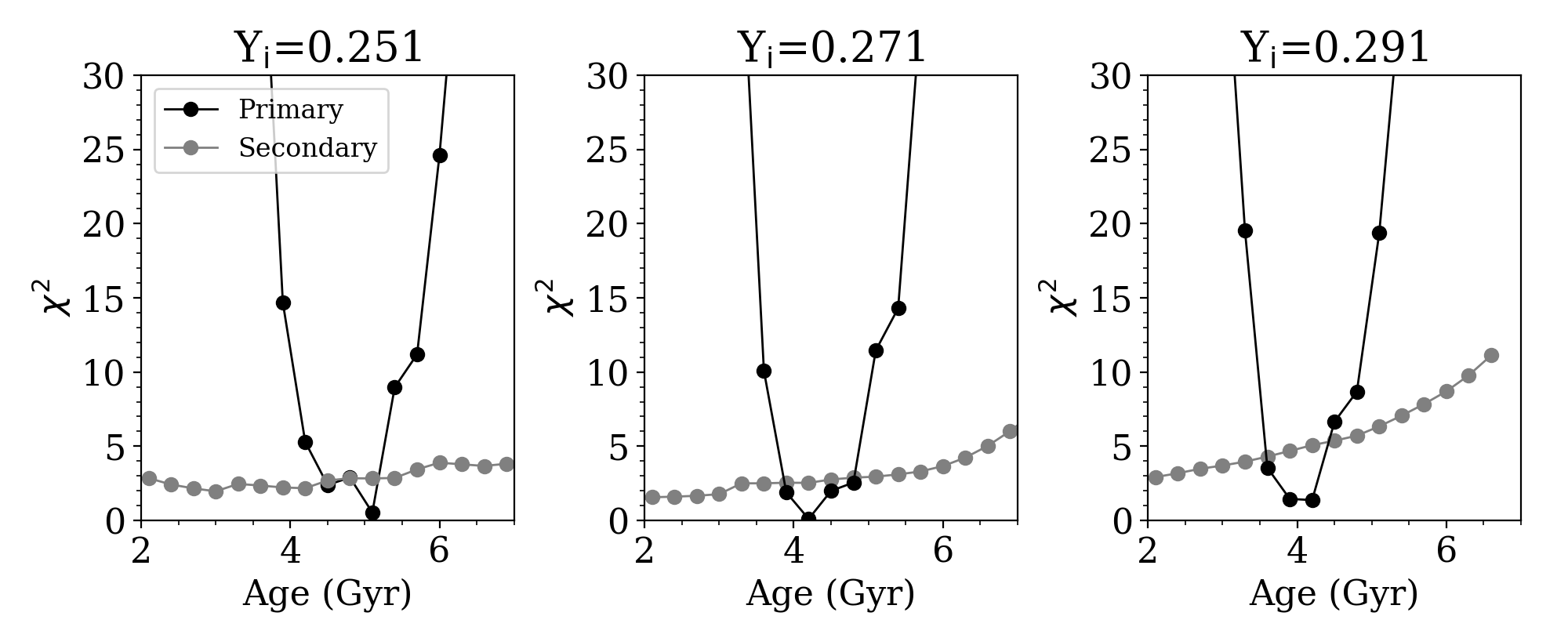}}
\caption{Variation of the overall fit to the observed parameters at different ages, for the primary (black) and secondary (grey) components. Each panel represents models calculated with a different initial helium abundance, which is given by the $Y_{\rm i}$ above the panel.} 
\label{fig:chisqAge}
\end{figure*}

Once the Bayesian analysis for each star and for each $Y_{\rm i}$ was complete, we used the best-fit age from the primary star's results to define the age of the system, and then found the matching age to get the parameters for the secondary. The total chi-squared, $\chi^{2}_{\rm tot}$, is calculated by summing $\chi^{2}_{1}$ and $\chi^{2}_{2}$ from the matching age slice. The best-fit parameters from the corresponding age-slices are shown in Table \ref{tab:AgeParams} are for stars of the same age and with the same initial helium abundance. The primary star was chosen to define the age of the system as its age was constrained more by the observed parameters. Figure \ref{fig:chisqAge} demonstrates this, as $\chi^{2}_{1}$ (solid points) forms a much narrower minimum compared with $\chi^{2}_{2}$ (light grey open points) from the fit to the secondary star's parameters. This is due the evolved nature the primary star. Stars will spend only a small fraction of their life on the subgiant branch, which in itself is a strong constraint on the age of a subgiant star, and the tight observational constraints limit it further.

\begin{table*}
\caption{{Best-fit evolutionary models for the primary and matching age model for the secondary, using different initial helium abundances $Y_{\rm i}$.}}

\label{tab:AgeParams}
\centering
\begin{tabular}{ l l |r r r r r r|r r}
\hline\hline
\noalign{\smallskip}
 \multicolumn{2}{c|}{Parameter} & \multicolumn{6}{c}{$Y_{\rm i}$} &\multicolumn{2}{|c}{Observed}\\
Symbol& Unit & 0.231 & 0.251 & 0.271 & 0.291 & 0.311 & 0.331 & Value & Error \\
\noalign{\smallskip}
\hline
\noalign{\smallskip}
$\tau_{\rm best}$ &{(Gyr)} 	& 5.31 & 5.05 & 4.22 & 4.12 & 3.55 & 2.98	& - & - 			\\
$T_{1}$ &(K)			& 6317 & 6297 & 6330 & 6340 & 6380 & 6356	&6330 & 50			\\
${\rm[Fe/H]_{{\rm s},1}}$ &-  	& $-0.47$ & $-0.35$ & $-0.35$ & $-0.23$ & $-0.23$ & $-0.22$&$-0.33$& 0.10\\
$M_{1}$ &$(M_{\sun})$		& 1.1539 & 1.1552 & 1.1568 & 1.1568 & 1.1495 & 1.1578	&1.1544& 0.0073	\\
$R_{1}$& $(R_{\sun})$		& 1.8324 & 1.8281 & 1.8331 & 1.8407 & 1.8503 & 1.7897	&1.834 & 0.023	\\
$\rho_{1}$&$(\rho_{\sun})$	& 0.1866 & 0.1882 & 0.1847 & 0.1847 & 0.1806 & 0.2010	 & 0.1873 &0.0071\\
$\log g_{1}$  &	-		& 3.97 	& 3.98 	& 3.98 	& 3.97 & 3.96 	& 4.00 & 3.974& 0.011	\\
${\alpha_{\rm ml_{1}}}$& -	& 2.043 & 2.048 & 1.704 & 1.921 & 1.639 & 1.501 & - &	- 		\\
${\chi^{2}_{1}}$	&-	& 1.96 & 0.53 & 0.10 & 1.37 & 3.31 & 5.39	&-&-			\\
\noalign{\smallskip}
\hline
\noalign{\smallskip}
$T_{2}$ &(K)			& 5363 & 5397 & 5410 & 5480 & 5487 & 5520	& 5400 & 	80		\\
${\rm[Fe/H]_{{\rm s},2}}$ &-	& $-0.53$ & $-0.45$ & $-0.35$ & $-0.26$ & $-0.16$ & $-0.13$&$-0.45$&0.11\\
$M_{2}$& $(M_{\sun})$		& 0.7784 & 0.7746 & 0.7766 & 0.7784 & 0.7795 & 0.7795& 0.7833&0.0050	\\
$R_{2}$& $(R_{\sun})$		& 0.7245 & 0.7251 & 0.7263 & 0.7317 & 0.7322 & 0.7339	&0.7291&0.0081\\
$\rho_{2}$& $(\rho_{\sun})$	& 2.041 & 2.027 &  2.021 & 1.982 & 1.979 & 1.966	& 2.023&0.073	\\
$\log g_{2}$	& -		&  4.61 & 4.61	 & 4.61    & 4.60   & 4.60     & 4.60  & 4.607 & 0.010	\\
${\alpha_{\rm ml_{2}}}$&-	& 1.810 & 1.864 & 1.880 & 1.962 & 1.973 & 1.754	&-&-		\\
${\chi^{2}_{2}}$	&-	& 1.64 & 2.83 & 2.53 & 5.07 & 8.89 & 11.67	&-&-			\\
\noalign{\smallskip}
\hline
\noalign{\smallskip}
${\chi^{2}_{\rm tot}}$	&-	& 3.60 & 3.36 & 2.63 & 6.44 & 12.20 & 17.06	&-&-			\\
\noalign{\smallskip} 
\hline
\end{tabular}

\end{table*}

Individually, the hotter primary star strongly favours an initial helium abundance of $Y_{\rm i} = 0.271$. A higher helium abundance means a higher mean molecular weight, which in turn means a higher luminosity \citep{1990sse..book.....K}, and therefore temperature, when $M$ and $\rho$ are fixed. However, with small uncertainties on the temperatures providing such tight constraints, it is the parameters with weaker constraints, such as $\alpha_{\rm ml}$ and [Fe/H] that have to adapt. From the solutions for the secondary star, the smallest $\chi^{2}$ is seen for $Y_{\rm i} =0.231$, which is lower than that of the primary component. This solution sacrifices the temperature in order to improve the fit to the mass and radius. This value is in fact less than the primordial helium abundance at the time of the big-bang nucleosynthesis, $Y_{\rm BBN}=0.2485$, \citep{2010JCAP...04..029S}. If this solution is excluded for this reason, $Y_{\rm i} = 0.271$ is the next best solution for the secondary star. This matches the value found for the primary.

After summing the $\chi^{2}$ from the individual stars, the best helium abundance for the system as a whole is 0.271, although it could be argued that a range between 0.231-0.271 could provide reasonable solutions, using $\Delta \chi^{2}_{\rm tot}=1$ as an estimate of a 68\% confidence interval \citep{1992nrfa.book.....P}. We refrain from quoting a true confidence region due the discontinuous nature of the helium abundances tested, and because there is no interpolation between the individual tracks. Better solutions could exist between the models. Interpolation has not been employed in these models due to the large number of parameters which would need to be considered. If the same estimate of a confidence interval for $\chi^{2}_{1}$ is used then the range of helium abundances is $Y_{\rm i} =0.251$-0.291, however $Y_{\rm i} =0.291$ is disfavoured by the secondary due to a high temperature and high metallicity.

\begin{figure}
\centering
\resizebox{\hsize}{!}{\includegraphics{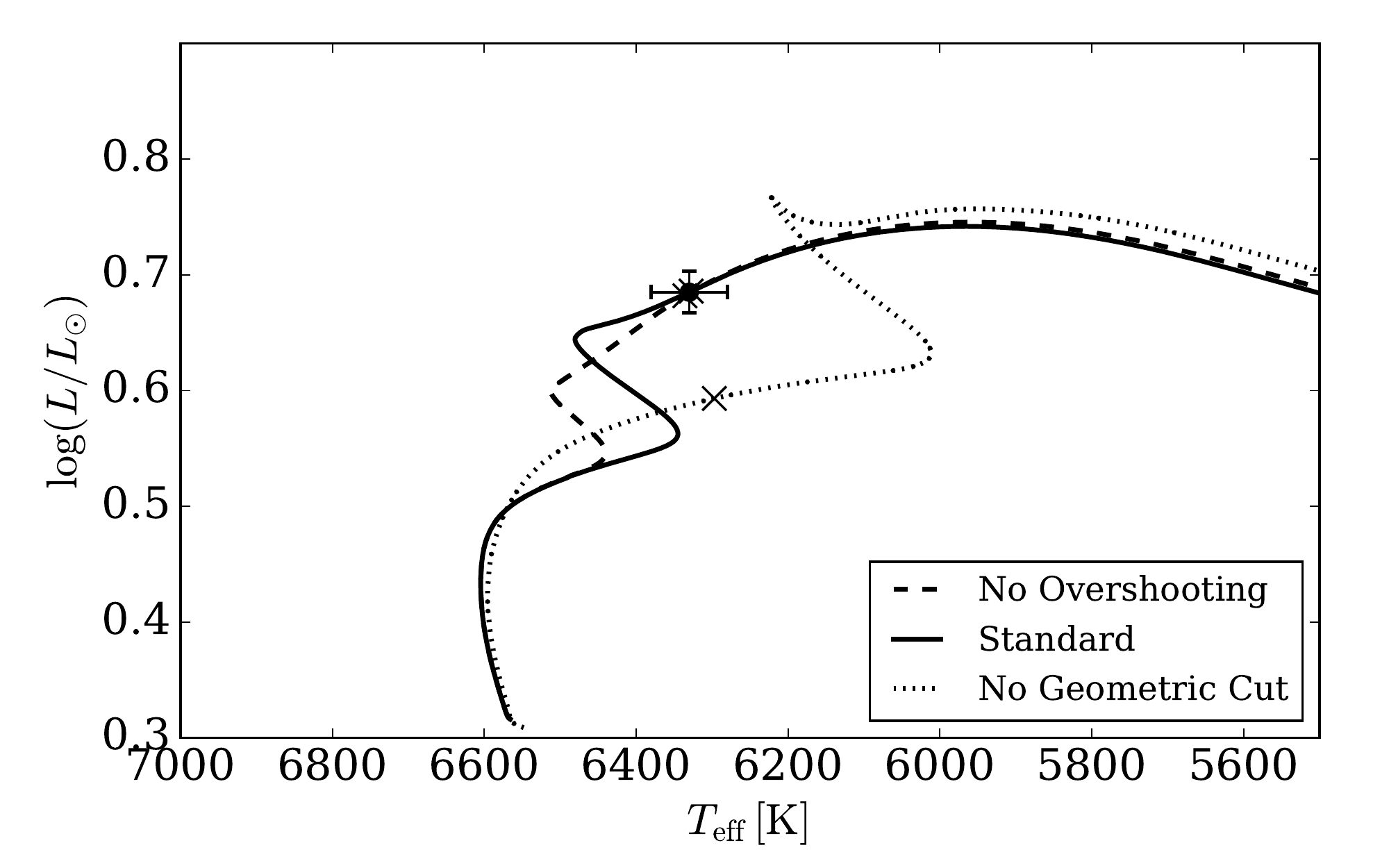}} 
\caption{ Evolutionary tracks in the temperature-luminosity plane for a 1.16\,M$_{\sun}$ star, with [Fe/H] =$-0.3$, $\alpha_{\rm ml}=1.70$ and $Y =0.271$, for three parameterisations of the overshooting parameter. Crosses mark the best-fit age obtained from the $Y=0.271$ fit in Table \ref{tab:AgeParams} for the primary star in WASP\,0639-32. All tracks are plotted from an age of 40\,Myr.} 
\label{fig:overshootPlot}
\end{figure}

The lowest $\chi^{2}_{\rm tot}$ suggests the age of WASP\,0639-32 is 4.22\,Gyr. Taking the uncertainty of $Y$ into consideration, as $\pm1$ in $\chi^{2}_{1}$, the overall age of the system is $4.2^{+0.8}_{-0.1}$\,Gyr. The $Y_{\rm i}=0.231$ solution was excluded as it is less than $Y_{\rm BBN}$. The age estimate does not consider the potential effects of different overshooting efficiencies, $f$, or different parameterisations of the overshooting.  The age of the primary star may be particularly sensitive to the overshooting, as its mass places the star on the boundary for developing convective cores \citep[1.1-1.2\,$M_{\sun}$,][]{2014EAS....65...99L}. A full investigation into the overshooting parameter is not possible with our current set of models, but there are a few simple tests to show how our choice of overshooting may affect the age estimate. For the tests, we consider a 1.16\,M$_{\sun}$ star, with [Fe/H] =$-0.3$ and $\alpha_{\rm ml}=1.70$. These are the best tracks for the Y=0.271 solution for the primary star. Figure \ref{fig:overshootPlot} shows the tracks for three different overshoot parameterisations. The track marked `standard' is the same parameterisation that is described in Sect. \ref{subsec:modsDescripts} and that has been used throughout the modelling so far, i.e. for a 1.16\,M$_{\sun}$ star where the geometric cut plays a role and the overshooting is partially suppressed. We also present the case where there is no overshooting and the case where the geometric cut does not play a role. The removal of the geometric cut means the overshooting is overestimated for the star, and provides an upper limit that is approximately equal to 0.3\,$H_{\rm p}$. 
Figure \ref{fig:overshootPlot} shows that removing the geometric cut results in a track that is a poor choice for the primary star. It also shows that reducing the level of overshooting will have little effect on the resulting age of the system. The same trends are seen for tracks with different helium abundances. Although, only one mass-track has been considered here, the small uncertainties on the observed masses would restrict the possible values for the overshooting. The effects of mass uncertainties on the overshooting parameter has been discussed in \cite{2017A&A...600A..41V}. It is worth noting that any uncertainty in the overshooting used for the primary star could affect the parameters that are presented for the secondary (as they are matched to the age of the primary star) and therefore $\chi^{2}_{\rm tot}$. However, as discussed above, the effect should be small.

For the lowest $\chi^{2}_{\rm tot}$ solution, the two stars use slightly different mixing-lengths, with the more evolved primary star requiring a value that is less than the solar value. The smaller secondary component prefers a mixing length that is about solar. This is consistent with the results from 3D radiative hydrodynamic models of \cite{2015A&A...573A..89M}. There is some degeneracy between $Y$ and $\alpha_{\rm ml}$ as they both affect the temperature, but the extent of the degeneracy is not fully understood. The degeneracy would need to be explored in more detail, for a strong claim to be made about the mixing lengths compared to those by \cite{2015A&A...573A..89M}, but this is beyond the capabilities of our current grid of models. One point to note about this work is the different solar compositions used between the spectroscopy \citep{2009ARA&A..47..481A} and these models \citep{1993oee..conf...15G}. This may introduce some discrepancies in the modelling of the temperatures, although by using a temperature that is obtained as a weighted mean from multiple methods, this discrepancy should be reduced. As the mixing length parameter has been included as a separate free parameter for both stars, it is possible that these free parameters are masking discrepancies between the models and observations, i.e. by picking a higher/lower mixing length it is possible to find a better match to the temperatures.

\begin{figure}
\centering
\resizebox{\hsize}{!}{\includegraphics{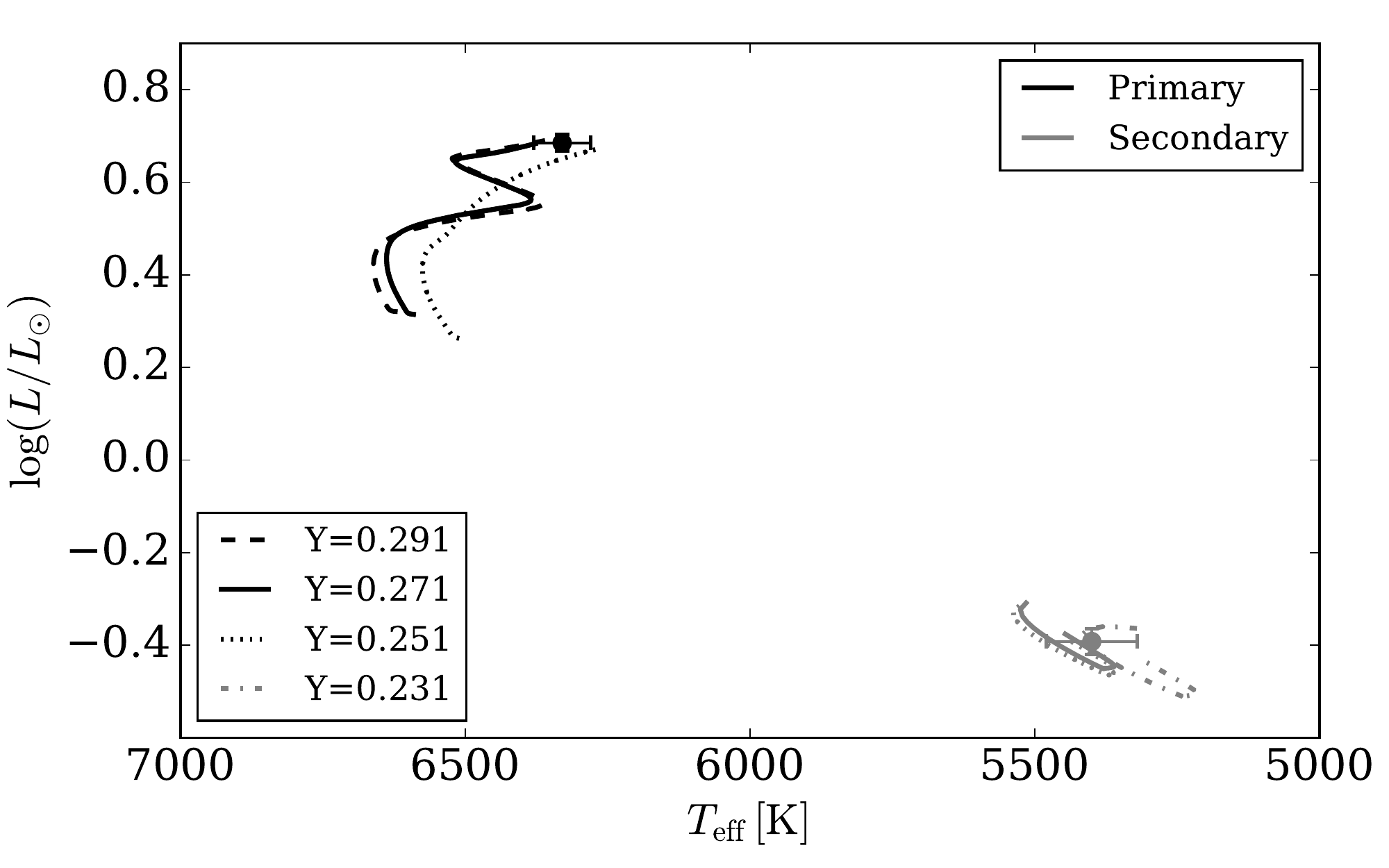}} 
\caption{Evolutionary tracks in the temperature-luminosity plane for the primary (black) and secondary (grey) components for different helium abundances. Dashed, $Y=0.291$; solid, $Y=0.271$; dotted, $Y=0.251$; dot-dashed, $Y=0.231$. All tracks are plotted from an age of 35\,Myr. Tracks closest the observed mass of each star are plotted. } 
\label{fig:lumModPlot}
\end{figure}

\begin{figure}
\centering
\resizebox{\hsize}{!}{\includegraphics{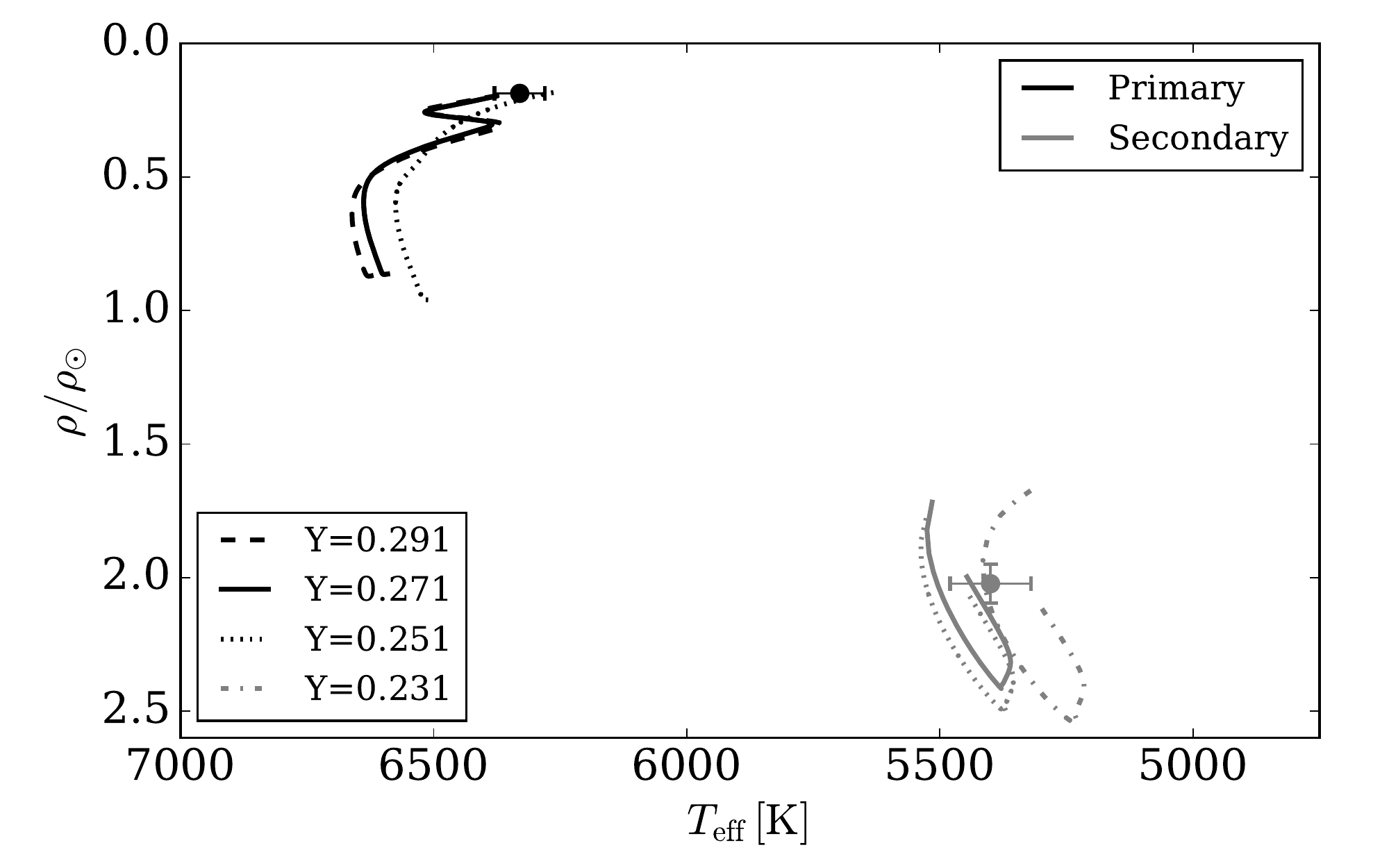}} 
\caption{Evolutionary tracks in the temperature-density plane for the primary (black) and secondary (grey) components for different helium abundances. Dashed, $Y=0.291$; solid, $Y=0.271$; dotted, $Y=0.251$; dot-dashed, $Y=0.231$. All tracks are plotted from an age of 35\,Myr. Tracks closest the observed mass of each star are plotted. } 
\label{fig:densModPlot}
\end{figure}

Figures \ref{fig:lumModPlot} and \ref{fig:densModPlot} show the best fit evolutionary tracks for the three best values of initial helium abundance, for each star. The points are plotted with the temperatures and luminosities from Table \ref{tab:MassRadiiData}.

\section{Discussion}
\label{sec:Discuss}

The trends in $\chi^{2}$ in Table \ref{tab:AgeParams} are quite clear for each star, making it possible to identify the preferred values for $Y_{\rm i}$. However, with the priors on the mass being looser than the observed value, there is a chance the fit to the mass has been sacrificed in order to provide a better match to other parameters such as the temperatures. A finer grid of models, particularly in regards to the mass and density parameters, would allow the tighter observation constraints to be used on these parameters and would enable the helium abundance to be constrained further by making the best $\chi^{2}$ value more distinct. As a test, the tracks were also fitted using the measured uncertainties on the the masses and densities, and this allowed $Y_{\rm i}$ to be constrained to the $Y_{\rm i}=0.271$ solution. However, the solutions presented very jagged $\chi^{2}$ minimisations plots due to the lack of interpolation in the models particularly for the secondary since the model step-size accounts for a larger percentage of the mass for the secondary component. Storing the required number of tracks to meet the precision of the measured masses and densities would be a huge challenge, with the current set of tracks reaching $\approx22\,000$ in number. An alternative would be to create a small subset of models around the solution obtained in this work, using a higher resolution. However, that is something that is beyond the scope of this paper. One approach taken for the binary system {\object{TZ For}}, was to fix the mass of the stars \citep{2017A&A...600A..41V} as the measured uncertainties were so small \citep[$\pm0.001\,M_{\sun}$,][] {2016A&A...586A..35G}. \cite{2017A&A...600A..41V} also state that this is not valid for uncertainties of the order $0.01\,M_{\sun}$. The uncertainties for the stars in WASP\,0639-32 lie between these two cases (at $0.0043\,M_{\sun}$), so it is currently unclear if fixing the masses of WASP\,0639-32 would hamper the determination of the helium abundance.

Initial attempts to determine the age of WASP\,0639-32 were done using the same evolutionary modelling method as was used for AI Phe \citep{2016A&A...591A.124K}. Here, both stars were fitted simultaneously with the effects of different mixing lengths being explored independently of the initial helium abundance. The lowest $\chi^{2}$ values occurred for mixing length that were solar ($\alpha_{\rm ml}=1.78$) and $\Delta Y =0.0$, and yielding $Y_{\rm i}=0.257$ with a range of $\pm0.020$. The two modelling methods use different chemical abundances meaning direct comparison should not be made, but it worth noting that this value matches the two initial helium abundances with the lowest $\chi^{2}_{\rm tot}$ in Table \ref{tab:AgeParams}.
The spectroscopic parameters obtained using a free $\log g$ were also tested using the method of \cite{2016A&A...591A.124K} to see if the evolution of the system differs. With the primary star set with a temperature of $6730$\,K, a good fit could not be found. The best solutions were suggesting a mixing length larger than 2.32, and a helium abundances larger than $0.31$. Even these solutions were very poor fits, with the models unable to match the densities and temperatures of both stars. This highlights the need to ensure accurate and consistent results for parameters that are used in stellar evolutionary modelling. It also highlights that caution should be used if the only source for a surface gravity is a spectroscopic value. \cite{2013A&A...558A.106M} shows that this issue is most prevalent at temperatures that deviates significantly from the Sun, but as yet, an exact cause has not been found. In \citeyear{2008A&A...488..653P}, \citeauthor{2008A&A...488..653P} looked at the atmospheric parameters for $\alpha$ Centauri A and B. They noted a disagreement between photometric and spectroscopic effective temperature scales for $\alpha$ Centauri A, with the spectroscopically derived temperature being higher than those obtained photometrically. This is similar to what has been found for the primary star of WASP\,0639-32. \cite{2008A&A...488..653P} suggest non-local thermodynamic equilibrium (NLTE) effects as a cause for the offset. As no NLTE effects are considered in the analysis in this paper, it is possible that this is the cause of the disagreement between the temperatures obtained with the fixed and free surface gravities. With further data releases to come from Gaia, it will be possible to obtain model independent effective temperature measurements using the angular diameter and integrated fluxes of the stars. 

\section{Conclusion}
\label{sec:Conc}

We have determined the masses and radii of the stars in a previously unstudied binary system, WASP\,0639-32, to the precision of 0.37\% and 0.36\% for $M_{1}$ and $M_{2}$ respectively and 1.27\% and 1.12\% for the $R_{1}$ and $R_{2}$, respectively. Taking the weighted mean of the temperature from each technique for each star (excluding spectroscopy with a free surface gravity), the effective temperatures are found to be $6330\pm 50$\,K and $5400\pm80$\,K for the primary and secondary, respectively.

With precise measurements of its fundamental parameters, this detached eclipsing binary system is a good choice for calibrating both main-sequence evolutionary models using the secondary star, but also subgiant models using the primary star. We have shown that it is possible to constrain the helium abundance of the system to a range 0.251-0.271, with the possibility of constraining it further with a finer grid of models.

\begin{acknowledgements}
Based on observations collected at the European Organisation for Astronomical Research in the Southern Hemisphere under ESO programme 094.D-0190(A). WASP-South is hosted by the South African Astronomical Observatory and we are grateful for their ongoing support and assistance. Funding for WASP comes from consortium universities and from the UK's Science and Technology Facilities Council. J.A.K.-K. is also funded by the UK's Science and Technology Facilities Council. A.M.S. acknowledges support from grants ESP2015-66134-R (MINECO). This research has made use NASA's Astrophysics Data System.
\end{acknowledgements}

\bibliographystyle{aa} 
\bibliography{j0639_paper}
\end{document}